\documentclass[traditabstract]{aa}
\usepackage{graphicx,epsfig}
\usepackage{multirow}
\usepackage{verbatim}
\usepackage[figuresright]{rotating}
\usepackage{txfonts}

\usepackage{natbib}

\bibpunct{(}{)}{;}{a}{}{,} 
\def\kms{km\,s$^{-1}$}
\def\msun{M$_{\odot}$}

\title{ The imprint of a symbiotic binary progenitor on the properties of Kepler's supernova remnant}

\author{A.~Chiotellis\inst{1}\thanks{email: a.chiotellis@astro-uu.nl}
\and K.M.~Schure \inst{1,2}
\and Jacco~Vink \inst{1}}
\institute{Astronomical Institute, Utrecht University,
             P.O. Box 80 000, NL-3508 TA Utrecht, The Netherlands
\and Department of Physics, University of Oxford, 
             Clarendon Laboratory, Parks Road, Oxford OX1 3PU, United Kingdom}

\begin{document}

\abstract{
We present a model for the Type Ia supernova remnant (SNR) of SN 1604, also known  as Kepler's 
SNR. We find that its main features can be explained by a progenitor model of 
a symbiotic binary consisting of a white dwarf and an AGB donor star with an initial mass of $4-5~\rm {M_{\odot}}$. The slow,
nitrogen rich wind emanating from the donor star   has partially been accreted by the white dwarf, but has also created a circumstellar bubble. Based on  observational evidence, we assume that the   system moves with a velocity of 250 \kms.  Due to the systemic motion the   interaction between the wind and the interstellar medium has resulted in the formation of a bow shock,
which   can explain the presence of a one-sided, nitrogen rich shell.
We present two-dimensional hydrodynamical simulations of both the shell formation and
the SNR evolution. The SNR simulations show good agreement with the
observed kinematic and morphological properties of Kepler's SNR.
Specifically, the model reproduces the observed expansion parameters ($m=V/(R/t$))
of $m \approx 0.35$ in the north and
$m \approx 0.6$ in the south of Kepler's SNR.
We discuss the variations among our hydrodynamical simulations in light of the observations,
and show that part of the blast wave may have traversed through the one-sided shell completely.
The simulations   suggest a distance to Kepler's SNR of 6~kpc, or  otherwise require that
SN 1604 was a sub-energetic Type Ia explosion. Finally, we discuss the possible implications
of our model for Type Ia supernovae and their remnants in general.
}


\maketitle

\section{Introduction}\label{Sec:Intro}

Type Ia supernovae (SNe Ia) 
are of prime interest for many areas in astrophysics. 
They are important cosmological standard candles,
because of their high, and well calibrated \citep{phillips92}
peak luminosities. This has led to the realization that 
we appear to live in a Universe whose expansion is accelerating 
\citep{perlmutter98,garnavich98}.
In addition, SNe Ia are major contributors to the chemical
enrichment of the Universe, as they are 
the principal source of iron peak elements.

There is a consensus that SNe Ia are the result
of thermonuclear explosions of CO white dwarfs (WDs) in binary systems  that approach the Chandrasekhar mass due to either accretion from a companion star (single degenerate scenario), or by the merging of two WDs (double degenerate scenario) \citep[see the reviews by][]{hillebrandt00, Livio00}. 
The double degenerate (DD) scenerario is favored by stellar population synthesis models 
\citep[e.g.][]{Yungelson2000,Ruiter09,Claeys11}, but the explosion mechanism itself is far from clear.
The main problem  is that WD mergers lead to an off-center ignition, which converts carbon and oxygen into oxygen, neon and magnesium. This results in an accretion induced collapse and the formation of a neutron star, rather than a thermonuclear explosion \citep[e.g.][]{Saio85,Mochkovitch90,nomoto91}. 

For the single degenerate (SD) scenario  the most puzzling problem is the nature of the progenitor binary systems.  The WD, in order to reach the Chandrasekhar limit, should accrete and burn the material from its companion star at a rate of around $10^{-7}$~M$_{\odot}$yr$^{-1}$ \citep{1982ApJ...253..798N}.  For lower accretion rates the  accumulated material is liable to unstable burning, giving rise to nova explosions that expel more mass than is accreted. On the other hand, high accretion rates lead to the expansion of the WD's photosphere to red giant's dimensions and, in the absence of stabilizing effects of strong accretion winds \citep{hachisu96,1999ApJ...522..487H}, the system will undergo a common envelope episode, which does not lead to a SD SNe Ia \citep{Iben1984}.  The finetuning needed to create a SN Ia explosion is in sharp contrast with the relatively high probability of $\sim 15$\%
for stars in the 3-8 $\rm M_{\odot}$ range to eventually explode as SN Ia
\citep[e.g.][]{mannucci06,2007A&A...465..345D,maoz08}.

Our understanding of SN Ia would benefit greatly if
we would identify progenitor systems directly or indirectly through
the imprints they have on the SN spectra, or the supernova remnants.
Several studies  have been conducted along these lines, but reach different
 conclusions about the origin of the SNe Ia \citep[e.g.][]{Hamuy2003,Panagia2006,Patat2007,Hayden2010}.

\begin{figure}[t]
	\centering
		\includegraphics[trim=10 30 10 40,clip=true,width=\columnwidth]{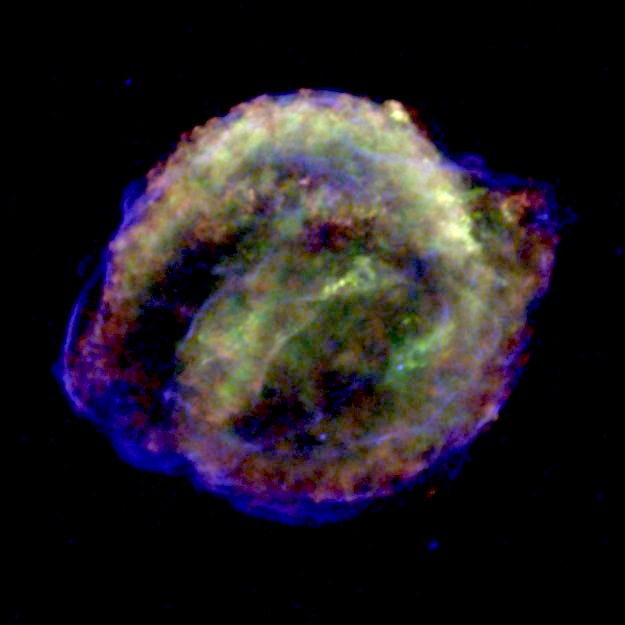}
	\caption{Chandra X-ray image of Kepler's SNR, with red indicating Si-K$\alpha$
emission (1.75-1.95 keV), green Fe-L emission (0.8-1.6 keV),
and blue continuum emission (4-6 keV). The image is based on a deep, 750 ks,
Chandra observation \citep{2007ApJ...668L.135R}.}
	\label{fig:chandra}
\end{figure}

Here, we report our investigation of what could be the origin of
the remarkable structure of the circumstellar medium (CSM) that shapes 
Kepler's SNR (Kepler for short), 
the remnant of the historical 
SN 1604 \citep{2003LNP...598....7G,2007ApJ...662..998B}. 
As we will discuss below, SN\,1604 was very likely a SN Ia and the CSM observed in the evolved SNR
puts constraints on the type  of progenitor of this SNR.

Kepler (G4.5+6.8) is located relatively
high above the Galactic plane, $590 d_5$~pc, with $d_5$ the distance
in units of 5~kpc. Its radius is 2.6$d_5$~pc. The distance itself is
not well known. \citet{1999AJ....118..926R} used the HI absorption feature to place a lower 
limit of $(4.8\pm1.4)$~kpc and independently placed an upper limit of 
$6.4$~kpc based on the lack of absorption by an HI cloud. \citet{2005AdSpR..35.1027S} combined an estimate of the shock speed based
on the H$\alpha$ line width with a proper motion measurement to derive
a distance of $d=3.9^{+1.4}_{-0.9}$~kpc.
 Although these
distance 
measurements agree with each other within the errors, the lack of a detection of
Kepler in TeV gamma-rays by H.E.S.S. \citep{2008A&A...488..219A},
coupled with gamma-ray model predictions \citep{2006A&A...452..217B}
and the energetics of
the SNR based on expansion measurements \citep{2008ApJ...689..231V}, suggest
a distance $\gtrsim 6$~kpc, or otherwise a subenergetic explosion.

The SNR shows bright optical nebulosity with prominent [NII] line emission
in the north, indicating dense material with elevated
nitrogen abundances, $[N]/[N]_\odot > 2$, but otherwise
roughly solar metallicity \citep{1991ApJ...366..484B}. 
Radio \citep{1988ApJ...330..254D}
and X-ray expansion measurements \citep{2008ApJ...689..231V,katsuda08} of the 
SNR show an overall expansion parameter 
$m =  V/(R/t)\approx 0.6$, with $V$ the plasma or shock velocity,
$R$ the corresponding radius, and $t$ the age of the SNR. An exception is 
the northern region, where the expansion parameter is $m\approx 0.35$, lower
than expected for the  Sedov--Taylor phase in a homogeneous medium ($m=0.4$).
Based on these results, \citet{2008ApJ...689..231V} estimates that the nitrogen-rich shell must
have had a mass of at least 1~$\rm M_{\odot}$ \citep[see also][]{2007ApJ...662..998B}.

The presence of the nitrogen-rich shell has puzzled astronomers for long and
has led to the claim that SN 1604 was a Type Ib SN \citep{1988LNP...316...81B}, 
with the shell being the shedded outer envelope of the progenitor.
In order to explain the height above the Galactic plane, the one-sided
morphology
of the shell and the high proper motion of the nitrogen-rich knots, \citet{1988LNP...316...81B} argued for a high proper motion of the
progenitor, $\sim 280$~km\,s$^{-1}$ \citep[see also][]{1992ApJ...400..222B}. This high systemic velocity of Kepler's progenitor was verified observationally based on the proper motion and radial velocities of the nitrogen-rich knots of the remnant \citep{1991ApJ...374..186B} and by  the $ \rm {H_{\alpha}}$ narrow component of the non-radiative shocks of the SNR, which gave $u_* \approx 250$~km\,s$^{-1}$  \citep{1991ApJ...366..484B,Sollerman2003} 

Since the late nineties it has become clear that SN 1604 was very
likely not a Type Ib, but a SN Ia. The main reason is that
the X-ray spectrum has prominent Fe-L emission  and relatively little oxygen emission \citep{1999PASJ...51..239K,2007ApJ...668L.135R},
characteristic of Type Ia SNRs \citep{hughes95}. Supporting evidence for a Type Ia identification
is the presence of Balmer dominated shocks 
and the absence of an X-ray emitting,
 cooling,  neutron star \citep{2007ApJ...668L.135R}.
This means that the dense circumstellar nitrogen-rich shell must originate
from the progenitor system of the SN Ia.

Here, we show that the characteristics of Kepler can best be explained
within the framework of a single degenerate SN Ia model, assuming
non-conservative mass transfer through wind accretion from a 
4-5 $\rm{M_{\odot}}$  
asymptotic giant branch (AGB) star. We retain in this model the 
idea that the northern shell is the result of a bow shock caused by
the  motion of the progenitor system \citep{1988LNP...316...81B,1992ApJ...400..222B}, and we adopt the observed systemic velocity of 250 \kms.
Hydrodynamical simulations show that this model can account for the morphology 
of the SNR and its expansion characteristics. 

In Sect.~\ref{Ia_scenario} we discuss the progenitor binary system and in Sect. \ref{sec:bubble_snr} we determine its implications on the requirements for the interstellar medium (ISM) and the evolution of the circumstellar medium and  SNR. 
In Sect.~\ref{sec:hydro} we model the system using hydrodynamic simulations with the appropriate parameters and discuss the differences when varying the exact parameters. We evaluate the results of the simulations and our progenitor model in a broader context in Sect.~\ref{Sect:Discuss.} and end with our conclusions in Sect.~\ref{sec:conclusions}.

\section{A type Ia progenitor scenario for SN 1604}\label{Ia_scenario}

A model of Kepler's progenitor system should explain the formation of a $ \geq 1~\rm{M_{\odot}}$ asymmetric shell with solar metallicity and enhanced nitrogen abundances, which lies at the northern region of the remnant at a distance of 2-3~pc from the explosion center. 

Potentially, this shell could have been formed by {\em i}): substantial outflows of the WD's surface such as nova explosions or `accretion winds', {\em ii}): the wind of the WD progenitor star (i.e. the SNR interacts with a planetary nebula-like shell),  {\em iii}): the wind of the donor star, or by {\em iv)}:  the ejected common envelope, in the case of a DD progenitor scenario.  

The nova explosions, which are related with SNe Ia are the reccurent novae that occur on the surface of massive WDs ( $\geq 1.2 \rm{M_{\odot}} $)  in the last phase of the binary evolution \citep{2001ApJ...558..323H,2008ApJ...679.1390H}. However, during the reccurent novae phase, the total mass that is ejected from the WD surface is of the order of $10^{-3} - 10^{-2} \rm{M_{\odot}}$ \citep{2008ApJ...679.1390H}. Given that recurrent nova ejecta have similar to solar abundances \citep[see][for summary]{Livio1992}, they are not able to accumulate enough heavy elements into the CSM shell to reproduce the observed chemical abundances. On the other hand,  an outflow in the form of an accretion wind emanating from the WD surface is so fast ($u_{wind} \sim 1000$ \kms) that it would form a large low-density cavity around the progenitor system. \citet{2007ApJ...662..472B} showed that these cavities are at odds with the observed radii, shock velocities  and ionization timescales of Galactic, LMC  and M31 Type Ia SNRs (among them  Kepler SNR was included). Our hydrodynamical simulation of a SNR evolution in a cavity formed by accretion winds verifies their results.

Case {\em ii} can be excluded because of the time scales involved. The time interval between the planetary nebula phase and the Type Ia exposion should be at least $\sim 10^6$ yr as the WD needs to accrete at least 0.2 \msun\ at a rate of $10^{-7}$ \msun\,yr$^{-1}$. In such a period the formed shell would have collapsed under the ram pressure of the interstellar medium as the binary system is moving with a velocity of 250 \kms.  The same line of argumentation can be applied for case {\em iv}, since, after the ejection of the common envelope the merging timescale of the two WDs is at least 0.1 Myr (Claeys 2011, private communication).  Although this scenario seems unlikely, the lack of understanding of the ejection of a common envelope and the subsequent evolution \citep[e.g.][]{Taam2010} prevents us from drawing a definitive conclusion about the likelihood of case {\em iv} as a progenitor scenario.

Therefore, the wind from the donor star (case {\em iii}) appears to be the most likely origin for this circumstellar shell as the wind velocities of evolved stars are much smaller than those of the WD's outflows, resulting in smaller and denser cavities and their formation sustains until the moment of the explosion.

\medskip\noindent
{\bf Chemical composition of the circumstellar shell}
 
The northern shell of Kepler  reveals a roughly solar metallicity plasma with elevated nitrogen abundances of at least 2 times the solar value. The only process to enrich the surface of a  low mass star with nitrogen is \textit{hot bottom burning} (HBB), which occurs in massive AGB stars \citep[$M>4~\rm{M_{\odot}}$, depending on metallicity,][]{2007PASA...24..103K}. During this process the convective envelope of the star dips into the outer layer of the H-shell, which results in a thin layer hot enough to sustain proton-capture nucleosynthesis. HBB converts $^{12}$C into $^{14}$N and if the third dredge-up is operating,  the stellar atmosphere will become nitrogen rich. Based on the results of \citet{2007PASA...24..103K},  the donor star of the progenitor system was most likely  an AGB star of $4-5 \rm{M_{\odot}}$ with solar metallicity (see Table \ref{tab:yields}).

\begin{table}
\centering
\begin{tabular}{c|c|c|c|c|c|c} 
 & \multicolumn{3} {c|} {Production Factor:} & \multicolumn{3}{c} {Wind's Abundances:} \\ 
  & \multicolumn{3} {c|} {$F=\log[<X_{\rm i,final}>/<X_{\rm i,initial}>]$} & \multicolumn{3}{c} {$[X_i]/[X_{i,\odot}]$} \\ \hline
             & $\rm ^{12}C$             &$\rm ^{16}O$               & $\rm ^{14}N$               &$\rm ^{12}C$ & $\rm ^{16}O$ & $\rm ^{14}N$ \\ \hline
$4~\rm M_{\odot}$ & $0.33$  & $-2.6\times 10^{-2}$  & $0.42$    & $2.1$   & 1.1        & 2.6 \\ \hline
$5~\rm M_{\odot}$ & $0.14$  & $-4.3\times 10^{-2}$  & $0.61$    & $1.4$   & 1.1        & 4.1 \\ \hline
$6~\rm M_{\odot}$ & $-0.22$ & $-8.1\times 10^{-2}$  & $0.91$    & $0.6$   & 0.8        & 8.1 \\ \hline
\end{tabular}
\caption{The production factor and the chemical abundances as a function of the stellar (initial) mass of an AGB star with solar metallicity \citep{2007PASA...24..103K}. For the case of $5~\rm{M_{\odot}}$ the AGB mass loss rates by \citet{1993ApJ...413..641V} have been used.} 
	\label{tab:yields}
\end{table}

\medskip\noindent
{\bf Symbiotic binary} \label{subsect:symb_bin}

Efficient mass accumulation on the WD requires stable mass transfer. Roche-lobe overflow (RLOF) by stars with deep convective envelopes, such as the AGB donor star of our scenario, leads to dynamically unstable mass transfer and the formation of a double WD binary system engulfed by their common envelope.  Such a system is not possible to  end up as a SD Type Ia SN.

Thus, the progenitor system requires a wide symbiotic binary, where, through non-conservative mass transfer, part of the donor's stellar wind is accreted onto the WD, while the rest is ejected from the system forming a circumstellar shell. Since a CO WD needs to accrete at least 0.2 \msun\ in order  to approach the Chandrasekhar mass, and the envelope mass of the AGB star is $\sim 3 -4$ \msun, an (average) accretion efficiency of $\geq 5 - 10\ \%$ is required for a Type Ia explosion. The mass transfer can progress by either pure wind accretion, or by the more promising and efficient method of wind-RLOF accretion as suggested by \citet{Mohamed2007}.  In either case, the AGB star should stay within its Roche lobe, implying that the orbital radius has to be larger than $1600~ \rm R_{\odot}$ for typical mass ratios.

\medskip\noindent
{\bf Systemic motion}\label{subsect. syst_motion} 

 Considering that the progenitor binary was moving with a systemic velocity of 250 ~\kms, Kepler' s Galactic latitude and the asymmetric accumulation of the circumstellar shell can be readily explained by its supersonic motion away from the Galactic plane \citep{1988LNP...316...81B,1992ApJ...400..222B}.  The interaction of the wind with the ram pressure of the interstellar medium causes the formation of a bow-shaped shell.  The subsequent supernova drives a blast wave that currently interacts only with the nearest part of this shell.

\section{Formation and evolution of the wind bubble  and of the subsequent SNR}\label{sec:bubble_snr}

Now that we have outlined the properties of the progenitor binary system, we can study the formation and evolution of the wind bubble and the subsequent interaction of the SNR with it. In this section, we approach the dynamics analytically and specify the variables, which determine our model.\\

\subsection{The formation of the wind bubble and the bow shock model}\label{subsect:bowshock}

The properties of a supersonically moving wind bubble, interacting with the ISM, is a function of four variables;  the mass loss rate, $\dot{M}$, the wind velocity $u_w$, the systemic  velocity,  $u_{\rm *}$, and the ISM density, $n_{\rm ism}$. The system will be in a steady state, when the timescale of the interaction process between the wind and the ISM will be larger than the flow timescale defined as: 

\begin{equation} \label{eq:tflow}
t_{\rm flow}\equiv \left(\frac{r}{u_w}\right)\approx 10^5 \left(\frac{r}{2~\rm{pc}}\right)\left(\frac{u_w}{15 \rm{~km~s^{-1}}}\right)^{-1} \rm{yr},
\end{equation}
where $r$ in this case indicates the distance of the bow shock from the mass-losing star \citep{1992ApJ...400..222B}. 

In a steady state, the position of the bow-shock shell is determined by the balance of the stellar wind and the ISM ram pressure. At the so-called stagnation point the wind termination shock  has the shortest distance, $r_0$, to the mass losing star. Equating the momentum fluxes of the ISM and the CSM, this can be estimated to be around: 
\begin{equation}\label{eq:stag}
r_0= 1.78\times10^3 \left( \frac{\dot{M_w}u_w}{n_{\rm ism} u^2_*} \right)^{1/2} \rm{pc},
\end{equation}
when the velocities are measured in $\rm km~s^{-1}$, $\dot{M}_w$ in $\rm M_{\odot}yr^{-1}$ and $n_{\rm ism}$ in $\rm cm^{-3}$ \citep{1982A&A...116..348H}.  

In Sect.~\ref{Ia_scenario} we argued that the donor star of the progenitor system was a $4-5~ \rm M_{\odot}$ AGB star. These stars are characterized by dense, slow, radiation-driven stellar winds with typical terminal velocities of $5-20$~km~s$^{-1}$  and temperatures of $1000-2000$~K \citep{2003agbs.conf.....H}. The mass loss rates are fluctuating in a wide range, starting from $10^{-8}-10^{-6} ~\rm M_{\odot}$yr$^{-1}$ during the early AGB phase to $10^{-6}-10^{-5} ~\rm M_{\odot}$yr$^{-1}$ at the thermal pulsating AGB phase \citep{1993ApJ...413..641V,2007PASA...24..103K}.  The total duration of the AGB phase is on the order of 1~Myr. 
Assuming typical wind loss parameters appropriate for the thermal pulsating AGB phase,
a steady state situation is reached after $t_{\rm flow}\approx 0.1 - 0.3$~Myr for $r_0= 2$~pc and $t_{\rm flow}\approx 0.1 - 0.4 $~Myr for $r_0= 3$~pc  (Eq.~\ref{eq:tflow}), which is well within the limits of the lifetime of an AGB star.
\\

 In order to reproduce the radius of the stagnation point, we also need an estimate of
the ISM density. We do not have a priori estimates of this density,
as there are no clear observational data for its properties in Kepler's neighborhood. Both the neutral/ionized warm and the hot ionized component extend to its latitude \citep{1991IAUS..144...67R}.   However, the constraint on the distance of the stagnation point ($r_0 \approxeq 2-3$~pc, see eq. \ref{eq:stag}) requires  $n_{ism}  \le  10^{-3} \rm cm^{-3}$. This value is characteristic for the hot ionized component ($T\approx 10^6$~K), where  $n_{ism}  \sim  10^{-4} - 10^{-3} \rm cm^{-3}$ \citep{McKee77}  and was also used by  \citet{1992ApJ...400..222B}, based on the observed densities of the wind shell 
\citep[$n_{shell} \approx 1- 10\ \rm cm^{-3}$,][]{1989ApJ...347..925S,1991ApJ...366..484B}.

\begin{figure}[tbp]
     \centering
\includegraphics[width=0.9\columnwidth,angle=0]{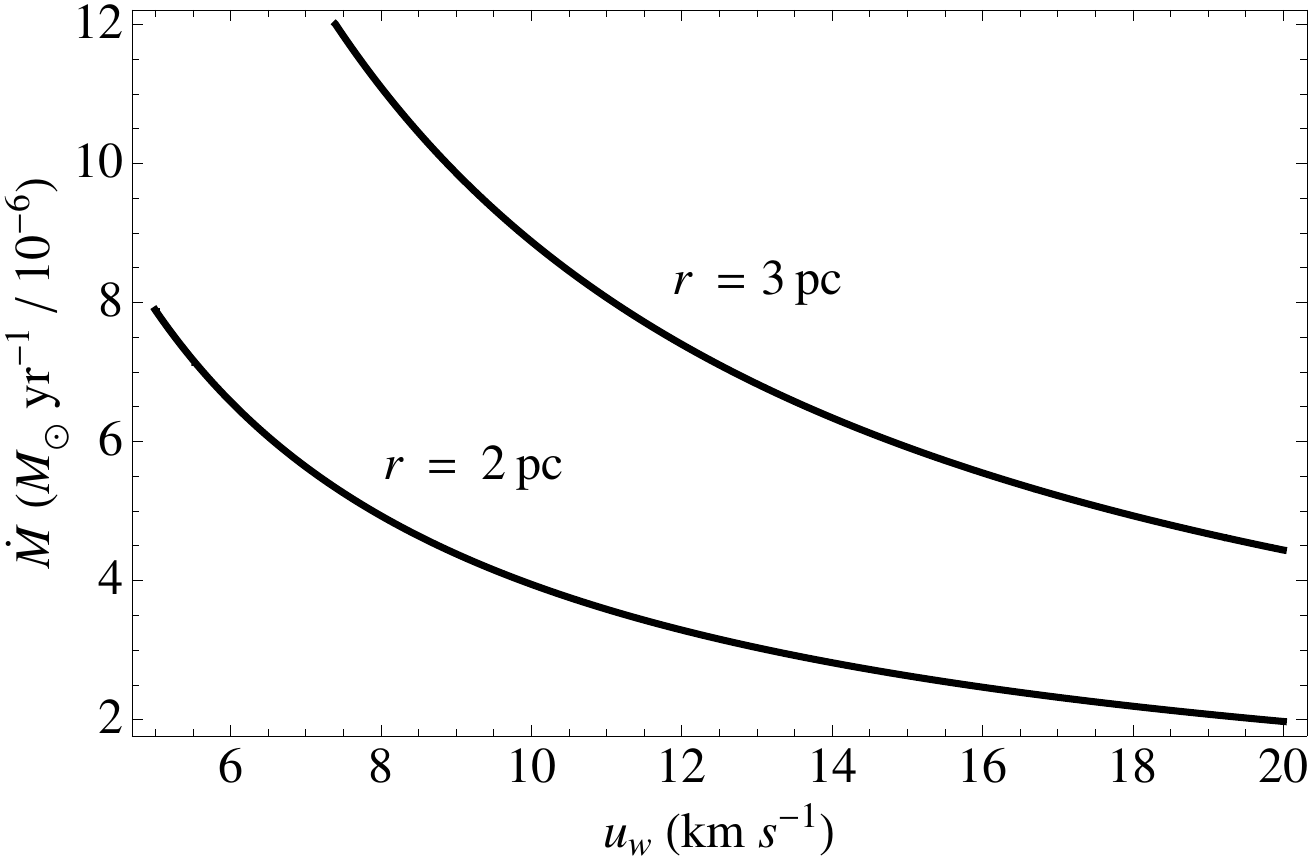} 
\caption{ The mass loss rates versus the wind velocity of the AGB star for the values which satisfy $r_0= 2$~pc and $r_0= 3$~pc for $n_{\rm{ism}}=0.5 \times 10^{-3}~\rm cm^{-3}$ and  $u_*= 250~ \rm{km~ s^{-1}}$.}
  \label{fig:stagnation} 
\end{figure}

Figure~\ref{fig:stagnation} shows the possible combinations of $\dot M_{\rm w}$ and $u_{\rm w}$ that result in $r_0= 2$~pc for the lower distance estimate of Kepler, and $r_0= 3$~pc for the larger distance estimate. In this plot we have adopted $n_{\rm{ism}}=0.5 \times 10^{-3}~\rm cm^{-3}$ for the hot ionized ISM, and  $u_*= 250~ \rm{km~ s^{-1}}$.  

\subsection{Supernova properties and evolution}\label{SN_prop}

The canonical values for the mass and energy of Type Ia SNe are $M_{\rm ej}=1.4$~M$_{\odot}$ and $E_{\rm ej}=10^{51}$~erg \citep{2007ApJ...662..487W}, respectively. We apply the self-similar solution of \citet{1982ApJ...258..790C} for the case where the SNR expands into an $\rho \propto r^{-2}$ wind profile, in our model corresponding to the regions interior to the shell. The ejecta is assumed to consist of a constant density core with an envelope that follows $\rho \propto r^{-7}$ density profile, while the velocity in the ejecta increases linearly. 
By assuming energy conservation, the expansion of the ejecta in a wind bubble with $\rho_{\rm wind} = q r^{-s}$ with $s=2$ and $q= \dot{M_w}/(4\pi u_w)$ is given by:

\begin{equation}\label{eq:rblast}
R_{\rm snr} = 1.3\times[A g^n/q]^{1/(n-s)}t^{(n-3)/(n-s)}   ,
\end{equation} 
where $A$ is a constant equal to $0.27$, and $t$ is the age of the remnant (in sec). Finally $g$ is a constant given by:
$$g^7 = (25/21\pi)(E_{\rm ej}^2/M_{\rm ej}) . $$

For the age of Kepler's SNR of $t\approx400$~yr, mass-loss parameters of $\dot{M_w} = 10^{-6} - 10^{-5} \rm M_{\odot}yr^{-1}$, and $u_w = 10 - 20$~km~s$^{-1}$, we find $R_{snr} = 3.0 - 4.7$~pc. These values of the SNR radius correspond to a distance of Kepler around $d = 6.2 - 9.7$~kpc. In order to obtain a SNR radius consistent with a distance of $d=4$~kpc,
we also have to consider a subenergetic explosion of $E= 0.2\times 10^{51} $~erg \citep[c.f.][]{2008ApJ...689..231V}. In that case we obtain $R_{\rm snr} = 1.6 -2.4$~pc and $d = 3.3 - 4.9$~kpc.

\section{Hydrodynamic modeling}\label{sec:hydro}

We employ the hydrodynamics code of the AMRVAC framework \citep{2003CoPhC.153..317K} for our simulations of the circumstellar bubble around Kepler's progenitor system and the subsequent evolution of the supernova ejecta. We perform the calculations on a 2D grid in spherical coordinates and assume symmetry in the third dimension. The Euler equations are solved conservatively with a TVDLF  scheme, using the adaptive mesh strategy to refine the grid where needed as a result of large gradients in density and / or energy. Our radial span is $2\times10^{19}$~cm and the range of the polar angle is  $0^\circ$ to $180^\circ$. On the base level, we use $240 \times 120$ cells ($R \times \theta$) and we allow for three refinement levels during wind evolution and four for the SNR evolution, at each of which the resolution is doubled. The maximum effective resolution, thus, becomes  $1.04 \times 10^{16}$~cm by $0.188^\circ$.  Radiative cooling is prescribed using the cooling curve by \citet{2009Schureetal}.

\begin{figure}[htbp]
	\centering
		\includegraphics[trim=80 20 320 80,clip=true,width=0.35\textwidth]{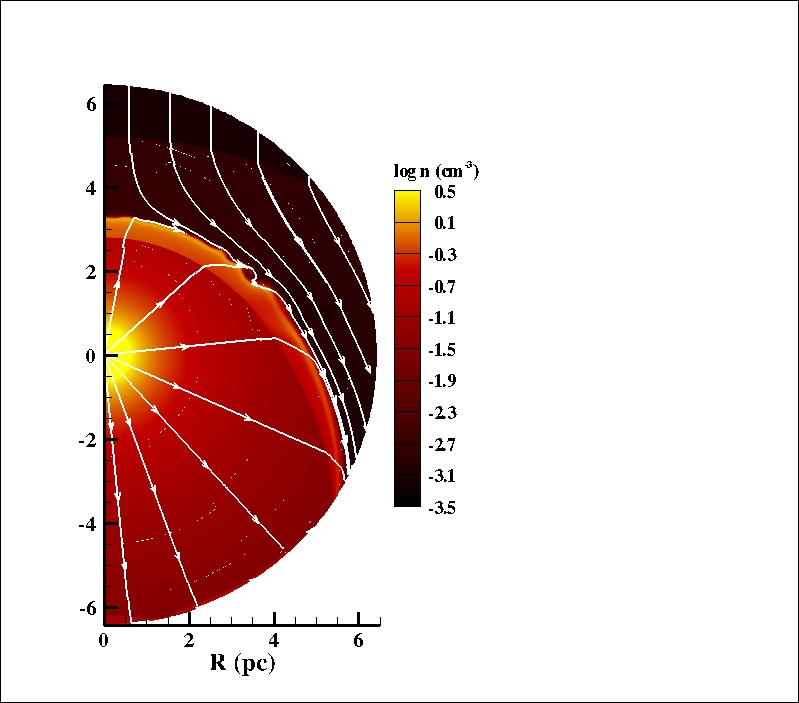}
	\caption{The 2D density profile of the simulation of the bow-shocked wind bubble. The mass-losing star is located in the origin where we let a radial wind flow enter the grid. Simultaneously, isotropic, homogeneous gas is entering the grid antiparallel with the y-axis, which represents the motion of the ISM in the star's rest frame. The arrows correspond to the vectors of the momentum of each flow. }
	\label{fig:orient_bubble}
\end{figure}

First we simulate the formation of the CSM bubble with the ISM bow shock shell. We model the system in the rest frame of the progenitor system and model the ISM interaction as an inflow. The ISM with density $\rho_i$ enters the grid antiparallel with the y-axis and with a momentum $m_r=\rho_i u_{*} \cos \theta$ (see Fig.~\ref{fig:orient_bubble}). Thus, the symmetry axis is aligned with the systemic direction of motion, roughly corresponding to  the northern region of Kepler's SNR. 
In the inner radial boundary, we impose an inflow in the form of a cold, slow, stellar
wind with a density profile of $\rho = \dot{M}_w/(4\pi r^2 u_w )$ and momentum $m_r = \rho u_w$ and $m_{\theta} =0$. 
In the second stage, we introduce the supernova ejecta into the wind bubble and allow the SNR to evolve.

In Sect.~\ref{sec:bubble_snr} we defined the constraints for the systemic velocity, mass-loss rate and wind velocity. These constraints
allow for some variation in those parameters, which will all give roughly similar values for the radius of the shell,
but which may nevertheless influence the properties of the SNR. In order to investigate this, we 
did several hydrodynamical simulations, varying the wind parameters and systemic velocity. In addition, as there
is some uncertainty concerning the distance to Kepler, we study the cases, where the stagnation point is placed at $\sim 2$~pc and at $\sim 3$~pc from the explosion center, corresponding to the SNR distances of 4~kpc and 6~kpc respectively. 
The grid of models is summarized in Table \ref{tab:5models}.

\begin{table*}[htbp]
	\centering
		\begin{tabular}{l |c|c|c|c|c|c }
		\hline
  	\textbf{Initial conditions}& \textbf{modelA}  & \textbf{modelB} & \textbf{modelC} & \textbf{modelD} & \textbf{modelDsub} \\
  		\hline
  	 $\dot{M}$~(M$_{\odot}$~yr$^{-1})$  &$10^{-5}$   &$7.5\times10^{-6}$ & $3\times10^{-6}$ &$7\times10^{-6}$ &$7\times10^{-6}$  \\
  	$u_{\rm w}$~(km~s$^{-1}$)          & 17 & 18 & 12 & 13 & 13 \\
         $u_{*}$~(km~s$^{-1}$)          & 250 & 250 & 240 & 250 & 250 \\
         $n_{\rm ism}~(cm^{-3})$  &$7\times 10^{-4}$&$5\times 10^{-4}$& $1.3 \times 10^{-4}$&$10^{-3}$&$10^{-3}$ \\  
        $t_{\rm bubble}$~(Myr)      & 0.38&0.52 & 0.85& 0.57& 0.57\\
       $E_{\rm sn}$~(erg)             & $10^{51}$& $10^{51}$& $10^{51}$& $10^{51}$& $0.2\times10^{51}$  \\
       
		\end{tabular}
	\caption{The properties of the 5 studied models. Additional parameters that have been used for all the models are:  $T_{\rm w} =1000~{\rm K}, T_{\rm ism} = 10^6$~K and $M_{\rm ej}= 1.4\ \rm{M_{\odot}}$. }
	\label{tab:5models}
\end{table*}

\subsection{Wind bubble evolution}\label{subsubsect:wbsn_evol}
\label{sec:bubbleA}

\begin{figure*}[htbp]
\begin{center}$
\begin{array}{cccc}
\includegraphics[trim=55 25 320 45,clip=true,width=45mm,angle=0]{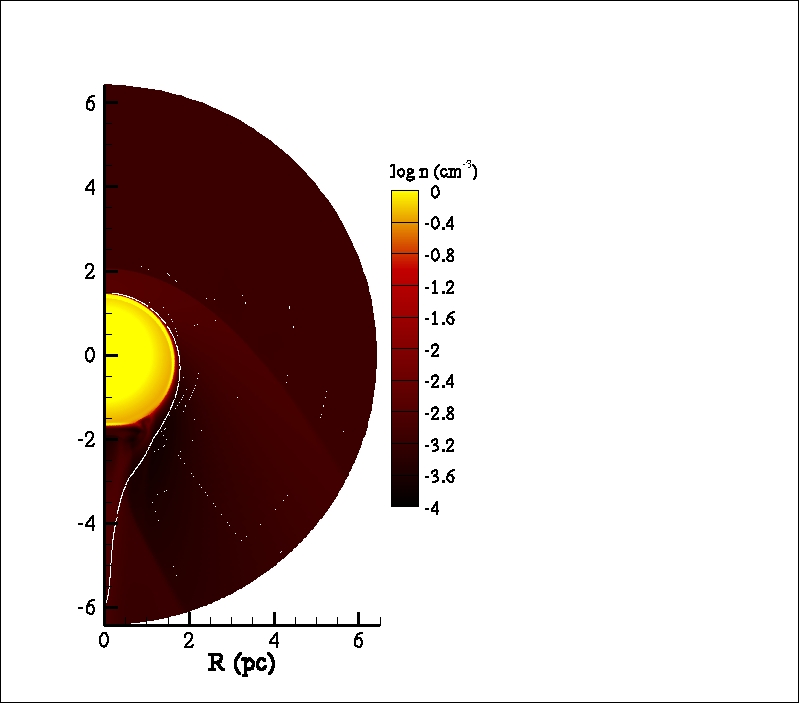} &
\includegraphics[trim=55 25 320 45,clip=true,width=45mm,angle=0]{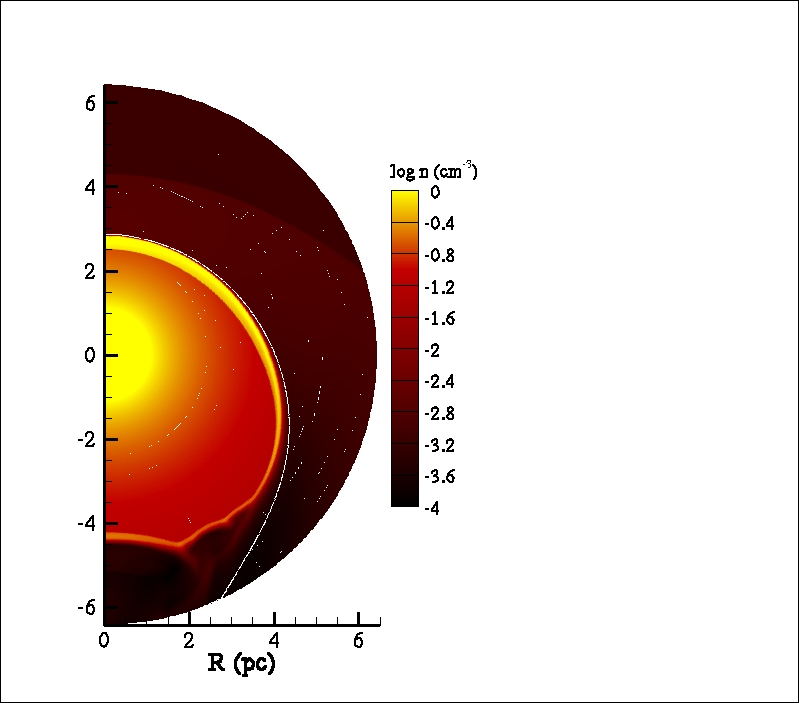} & 
\includegraphics[trim=55 25 320 45,clip=true,width=45mm,angle=0]{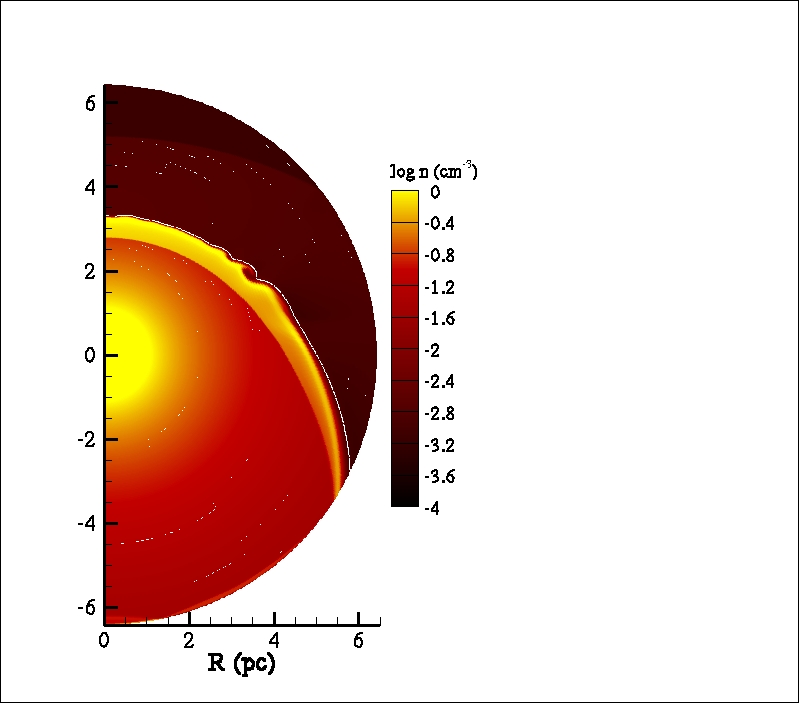} & 
\includegraphics[trim=55 25 320 45,clip=true,width=45mm,angle=0]{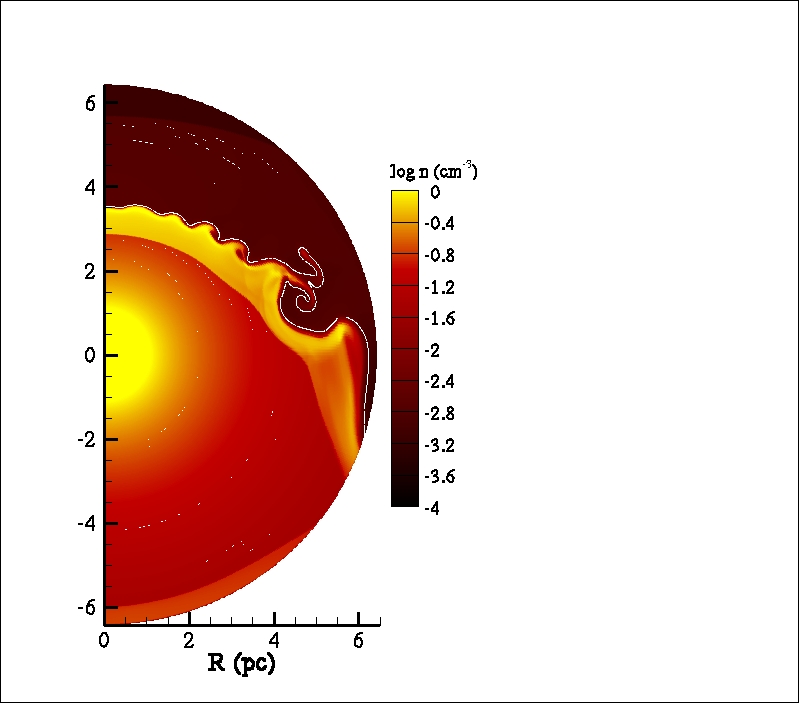} 
\end{array}$
\end{center}
\caption{The evolution of the wind bubble of model A. The snapshots from left to  right correspond to the times 0.10 Myr, 0.29 Myr, 0.38 Myr and 0.57 Myr. }
  \label{bubble_evol}
\end{figure*}

Figure \ref{bubble_evol} illustrates the evolution of the pre-supernova wind bubble using model A (see Table \ref{tab:5models}). It shows 
 the typical four-zone structure of a stellar wind bubble, consisting, from inside out, of the freely streaming wind with $\rho \propto r^{-2}$, the shell of shocked stellar wind, the shell of shocked ISM, and the unperturbed interstellar gas. At the border of each `zone' a density jump occurs due to the termination shock, the contact discontinuity, and the outer shock. The shear flow at the interface of the two fluids is susceptible to the Kelvin-Helmholtz instability, which results in a wavy structure  of the contact discontinuity.

The wind termination shock initially propagates rapidly outwards, but then decelerates, until it is stabilized, due to the momentum equilibrium between the wind and the ISM. This equilibrium is first achieved at the stagnation point, and then propagates to larger azimuthal angles. 

We introduce the SN explosion once the stagnation point is at 2-3 pc, while taking care so that the donor star's mass loss is less than the mass of the envelope of the AGB star and the duration is less than the life time of an AGB phase. 
In model~A, we find that these conditions are met at a time $t=0.38$~Myr after initializing the evolution of the bubble. The stagnation point in this case is located at a distance of  $\sim2.7$~pc, the outer shock is at a distance of $5.2$~pc, and the donor's mass loss is equal to $\dot{M} t= 3.8~\rm{M_{\odot}}$.

\begin{figure*}[htbp]
\begin{center}$
\begin{array}{cccc}
\includegraphics[trim=55 25 320 45,clip=true,width=45mm,angle=0]{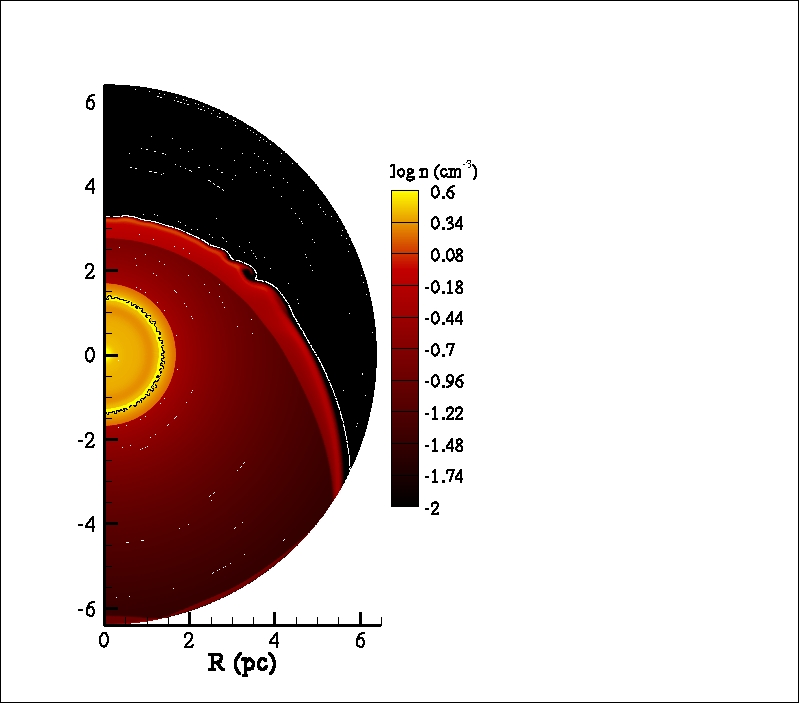} &
\includegraphics[trim=55 25 320 45,clip=true,width=45mm,angle=0]{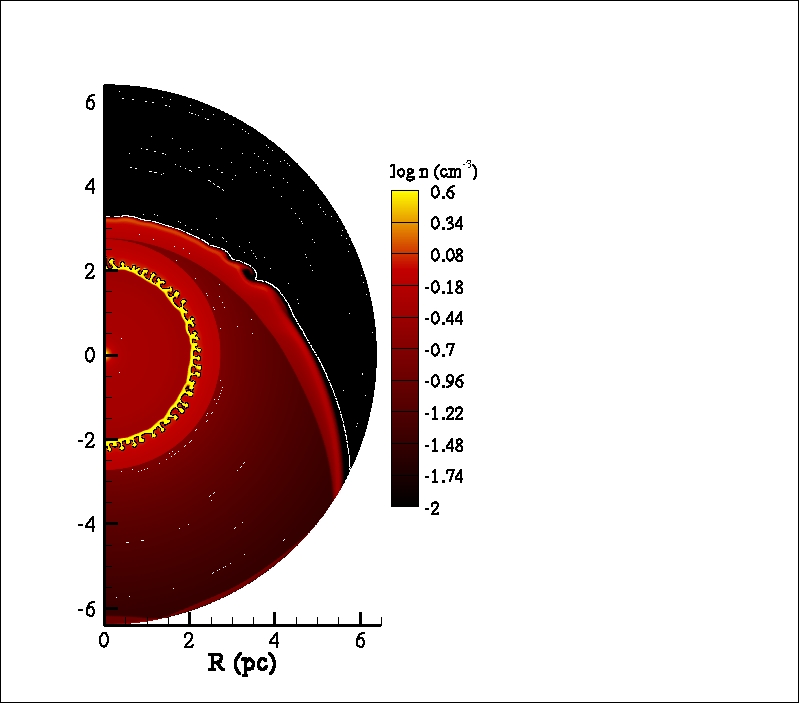}  & 
\includegraphics[trim=55 25 320 45,clip=true,width=45mm,angle=0]{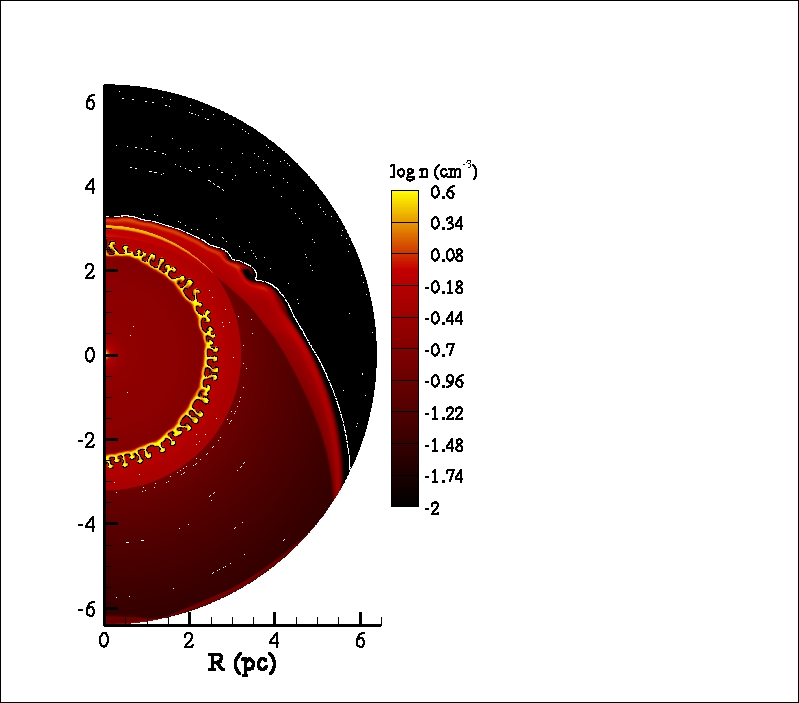} &
\includegraphics[trim=55 25 320 45,clip=true,width=45mm,angle=0]{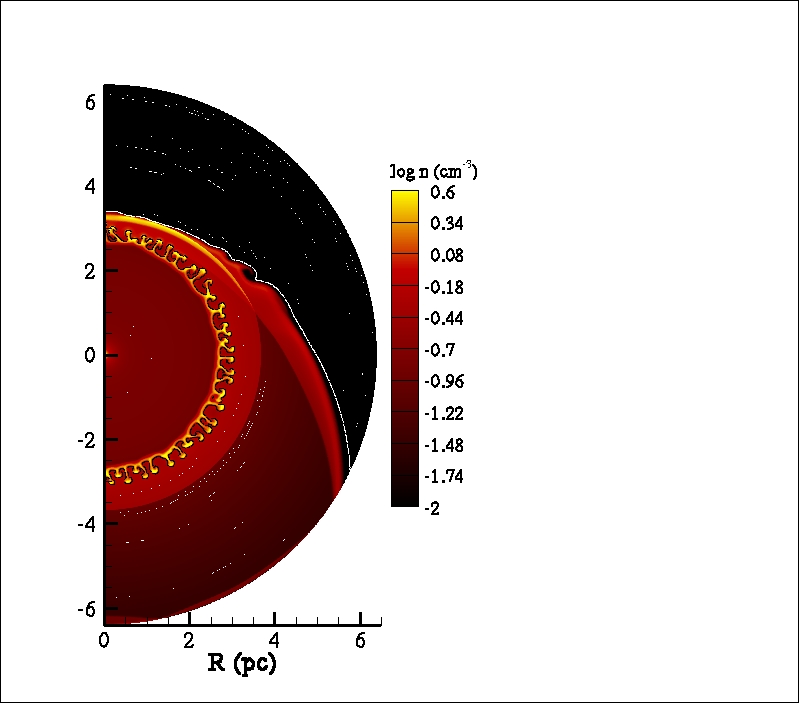} 
\end{array}$
\end{center}
\caption{SNR evolution of model A. The snapshots from left to  right  correspond to the times 158 yr,  285 yr, 349 yr and 412 yr. }
  \label{SNR_evol}
\end{figure*}

 \subsection{SNR evolution}
\label{sec:snrA}

Fig. \ref{SNR_evol} shows the evolution of the SNR for model A in an ambient medium that corresponds to the  wind bubble at time $t=0.38$~Myr.
Initially, the SNR  evolves in the spherically symmetric, freely expanding, wind-blown bubble (left panel). The freely streaming ejecta, shocked ejecta, contact discontinuity and the shell of the shocked CSM can be distinguished. Rayleigh-Taylor instability develops at the contact discontinuity between the ejecta and the shocked CSM. The thin black lines outline the interaction regions between  the ejecta and the CSM, while the white ones indicate the interface between the CSM and the ISM.
Around $\sim 285$~yr after the explosion (second panel) the blast wave starts to interact with the shocked stellar wind of the circumstellar shell in the area of the stagnation point. The blast wave sweeps up the dense shell's material and the deceleration in this area is therefore stronger, something that breaks the spherical symmetry of the SNR. 

With time, a larger portion of the SNR starts to interact with the circumstellar shell.  In the snapshot that corresponds to the age of Kepler's SNR ($t=412$~yr, right panel in Fig.~\ref{SNR_evol}) more than one third of the remnant interacts with the wind shell. Around the stagnation point, the blast wave has slightly penetrated the shell. At this moment, the blast wave in the southern region is located at a distance of $R_{\rm blast} \approx 3.6$~pc, and the contact discontinuity lies at $R_{\rm cd} \approx2.7$~pc. 
The highest densities are found in the area around the stagnation point, where it reaches values of $n\approx 4.0 ~\rm cm^{-3}$. This area is expected to have the highest emissivity. 

\begin{figure*}[htbp]
\begin{center}$
\begin{array}{cccc}
\includegraphics[trim=55 25 320 45,clip=true,width=45mm,angle=0]{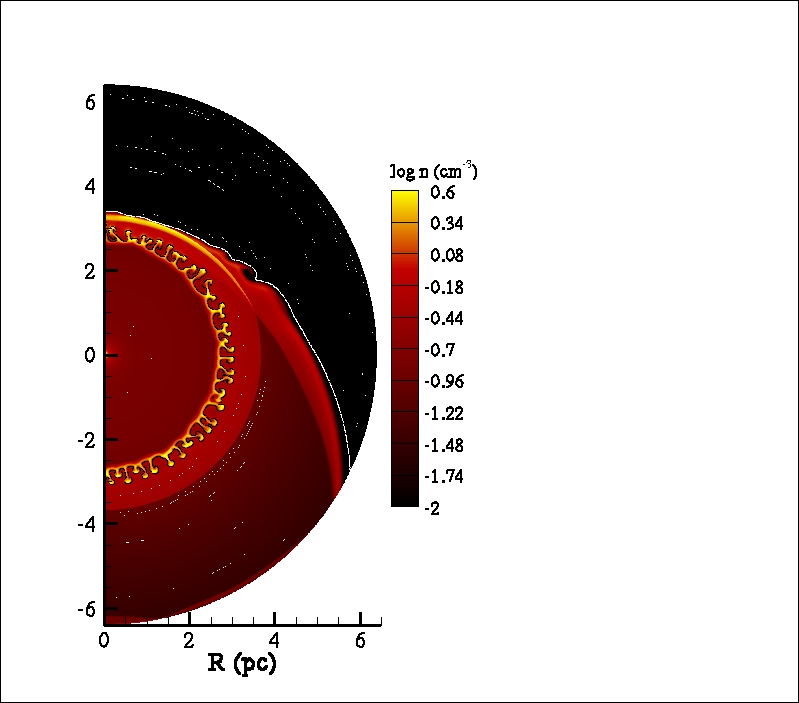} &
\includegraphics[trim=55 25 320 45,clip=true,width=45mm,angle=0]{figures_astroPH/Fig6a} &
\includegraphics[trim=55 25 320 45,clip=true,width=45mm,angle=0]{figures_astroPH/Fig6a} & 
\includegraphics[trim=55 25 320 45,clip=true,width=45mm,angle=0]{figures_astroPH/Fig6a}  \\
\end{array}$
\end{center}
\caption{Model A SNR properties at 412 yr. From left to right: density, expansion rate, pressure and temperature. }
  \label{SNR_A_prop}
\end{figure*}

Fig.~\ref{SNR_A_prop} shows the density, expansion rate, pressure, and temperature of the simulated SNR at the age of Kepler. The expansion rate, as expected, is much lower in the region, where the interaction with the circumstellar shell takes place. Around the stagnation, point we find values as low as $\sim0.35$, while for the non-interacting areas, the expansion rate is $0.6$. These values correspond  to the expansion rate of the plasma  and are consistent with results from X-ray observations \citep{2008ApJ...689..231V}.  The third panel shows the pressure,  which clearly marks the shocked plasma being rather homogeneous in the entire SNR, consistent with the expectation of pressure equilibrium between the shocks. The stronger interaction in the region around the stagnation point causes the SNR to be thinner and to have higher pressure compared to the rest of the remnant. 
The temperature plot (right panel) is limited to the range $\log T=6.0-9.5$,
corresponding only to the shock-heated SNR shell.
Note that X-ray observations usually provide electron temperatures,
which may be lower than the overall plasma temperatures depicted here due to
non-equilibration of ions and electrons.

\begin{figure}[hbt]
\centering
\includegraphics[trim=0 0 0 0,clip=true,width=85mm,angle=0]{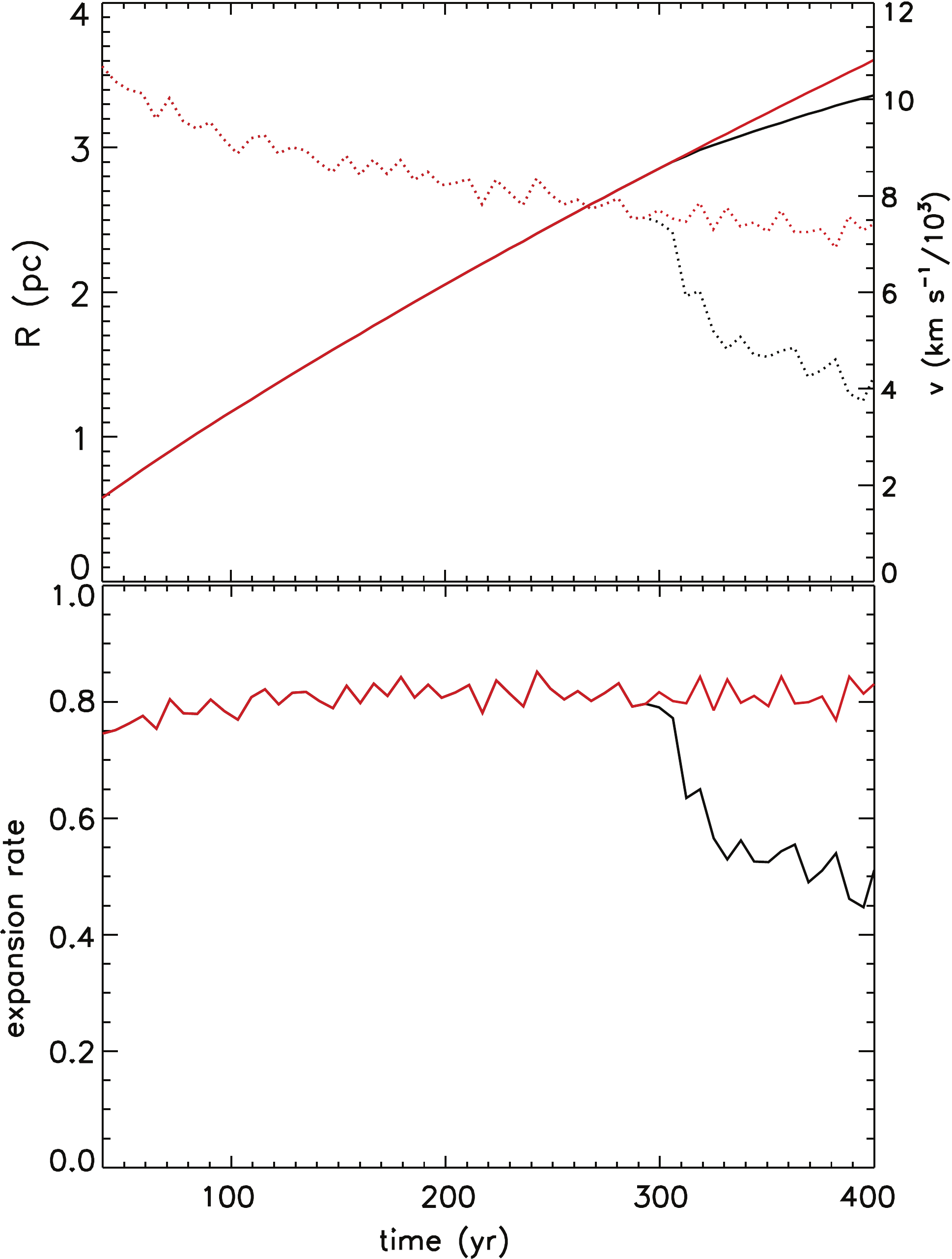}  \caption{Model A blast wave properties. The upper plot shows the position (solid lines) and the velocity (dotted lines) of the shock versus the time. The lower plot displays the evolution of the expansion rate. The black color corresponds to the azimuthal angle of $ 0^\circ$ while the red for $180^\circ$. }
  \label{Shock_x_v_exp}
\end{figure}

Fig.~\ref{Shock_x_v_exp} shows the position and the velocity of the blast wave (upper plot) and its expansion rate (lower plot) for two different azimuthal angles of $\theta = 0^\circ$ and $ \theta = 180^\circ$. During the evolution of the SNR in the freely-expanding wind region, these three quantities are identical for the two angles.  After 300~yr the collision with the shell starts and at the interacting region  ($\theta = 0^\circ$) the shock wave is strongly decelerated, while the SNR becomes slightly aspherical. The expansion rate is $\sim 0.8$ for the areas, which are still within the unshocked-wind region, while  for $\theta = 0^\circ$ it drops to $\sim 0.5$ for the SNR at Kepler's age.

\subsection{Comparison of the different models}\label{subsubsect:models}

In this section, we evaluate differences in the morphology of the SNR, when varying the mass loss properties and the systemic velocity of the system, as summarized in Table~\ref{tab:5models}.  

\begin{sidewaysfigure*} $
\centering $
\begin{center}$
\begin{array}{cccc}
\vspace{-0cm}
\includegraphics[trim=55 25 320 45,clip=true,width=45mm,angle=0]{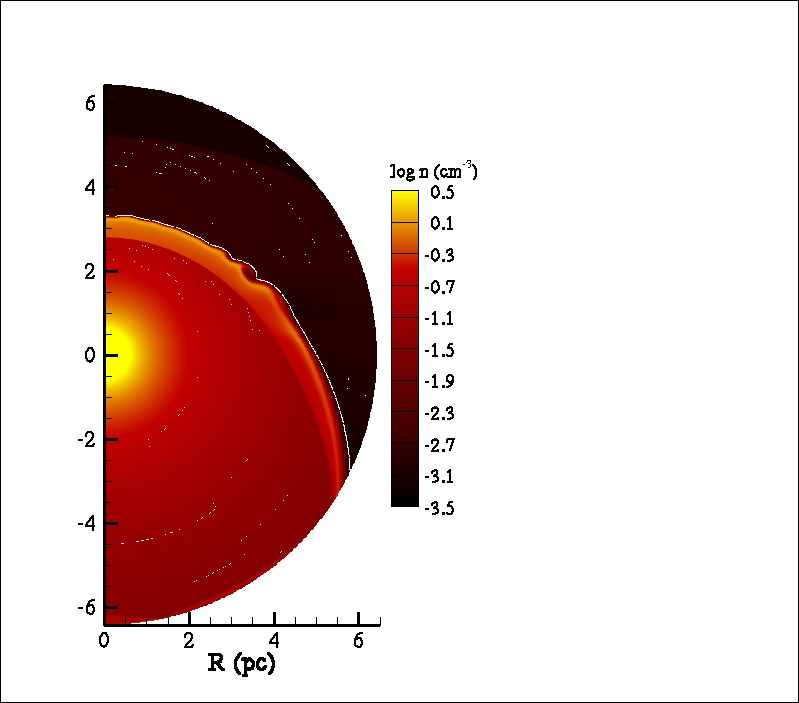} & 
\includegraphics[trim=55 25 320 45,clip=true,width=45mm,angle=0]{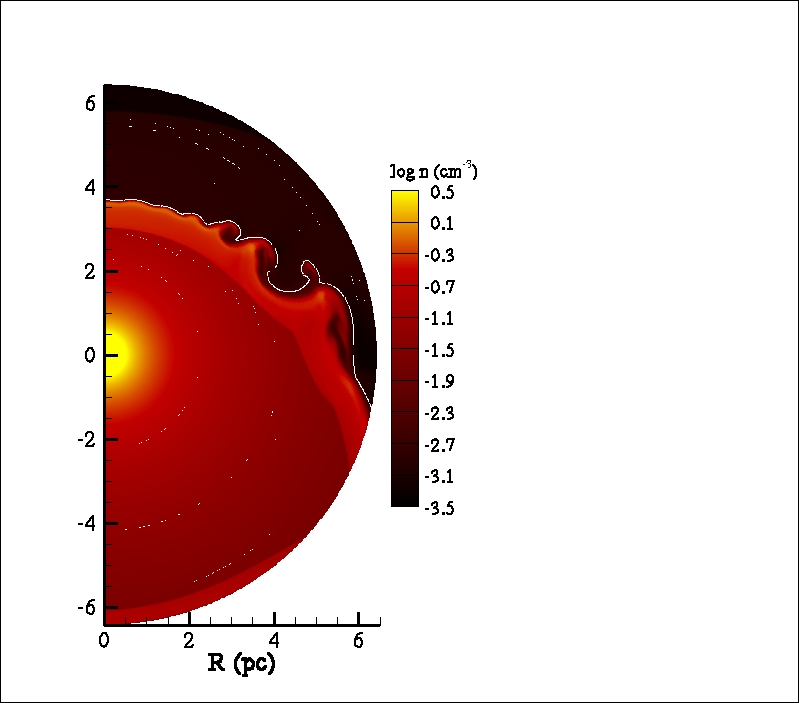} & 
\includegraphics[trim=55 25 320 45,clip=true,width=45mm,angle=0]{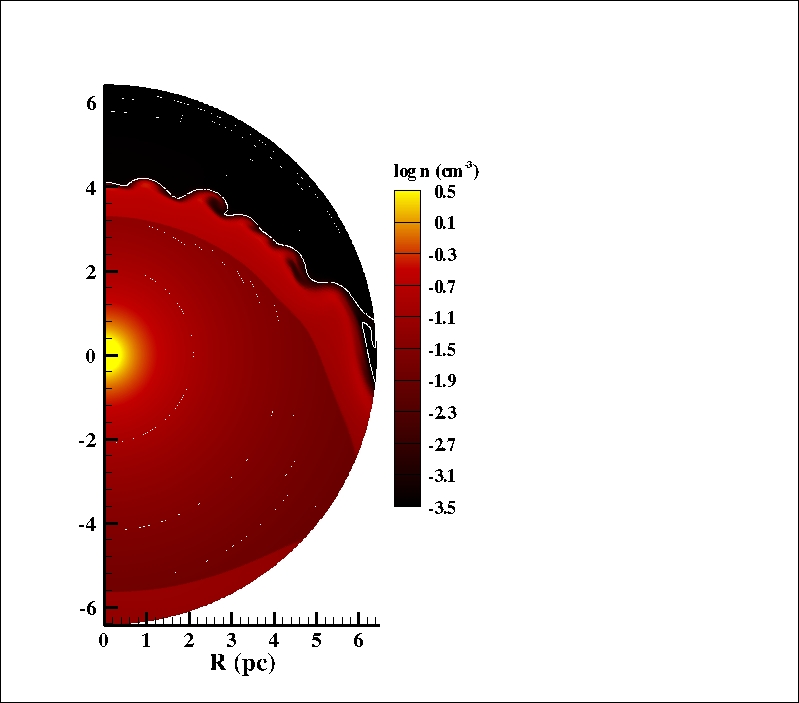} &
\includegraphics[trim=55 25 320 45,clip=true,width=45mm,angle=0]{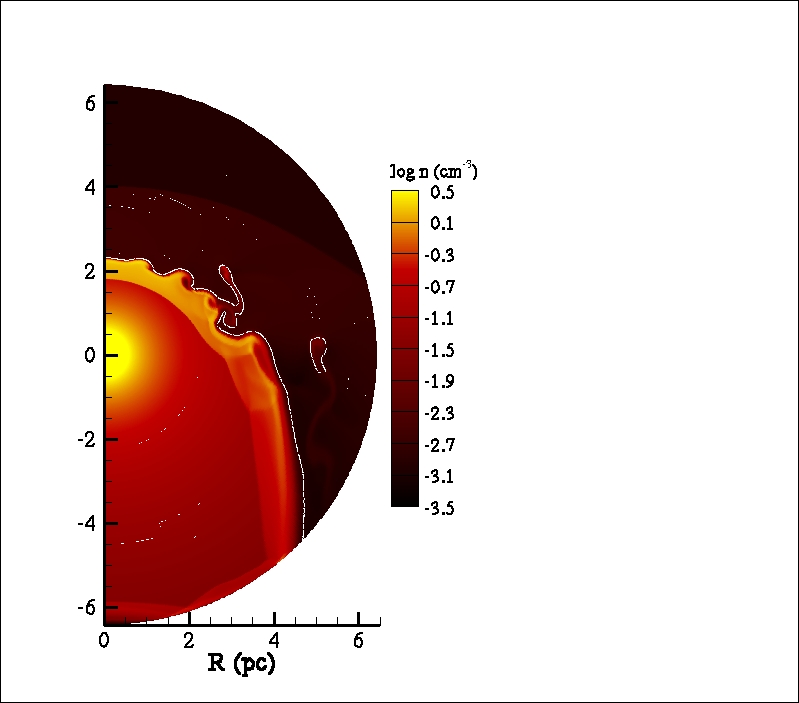}  
\end{array}$
\end{center}
\caption{Comparison of the four models of the wind shells. The plots from left to right  correspond to model A, B, C, D   (the bubble for model Dsub is the same as the one for model D).}
\vspace{-15cm}
  \label{bubble_comp}
\end{sidewaysfigure*}

\medskip\noindent
{\bf Wind bubble}\\
In Fig.~\ref{bubble_comp}, we show the densities of the bubbles for the different models, at the moment immediately prior to the SN explosion. The left panel shows model~A, which was already described in section~\ref{sec:bubbleA}. 

In model B we use a slightly lower mass loss rate for the formation of the wind bubble.  Due to the  longer duration of the shell formation, the Kelvin-Helmoltz instability at the contact discontinuity has developed more prominently. The chosen wind velocity is slightly higher, while the ISM density is lower in order to get the stagnation point at a distance of $\sim 3$~pc.  The wind termination shock and the contact discontinuity are farther from the star than in model A. This model has been considered in order to retain the blast wave of the SNR within the shocked wind shell at the current age of Kepler's SNR.

For model C we use an even lower mass loss rate and velocity of the AGB wind. These values fit better with the Reimers model for AGB mass loss parameters \citep{1975Reimers}, or an AGB at the early phase, according to the  \citet{1993ApJ...413..641V} description. In order to keep the stagnation point at a distance of 3~pc, also the systemic velocity and the interstellar medium density have been decreased. We consider a longer duration of the shell formation/mass transfer phase  in order to accumulate enough mass and  let the shell evolve to a distance of $\sim 3$~pc. The lower mass loss rates and wind velocities yield a lower momentum in the wind material, resulting in a weaker interaction and thus a thicker and comparitively more tenuous shell. 

The final panel shows the bubble for the model D.  The wind mass loss rate and duration of the shell formation are similar to those of model B. However, a lower wind velocity and a higher ISM density is applied. This causes the shell to be closer to the binary and denser.  This shell model has been considered in order to be able to include a case, where Kepler is located at a distance of $4-5$ kpc, for which we need the shell at a radial distance of $2.0 - 2.6$ pc.

\begin{sidewaysfigure*}
\begin{center}$
\begin{array}{ccccc}
\includegraphics[trim=55 25 320 45,clip=true,width=45mm,angle=0]{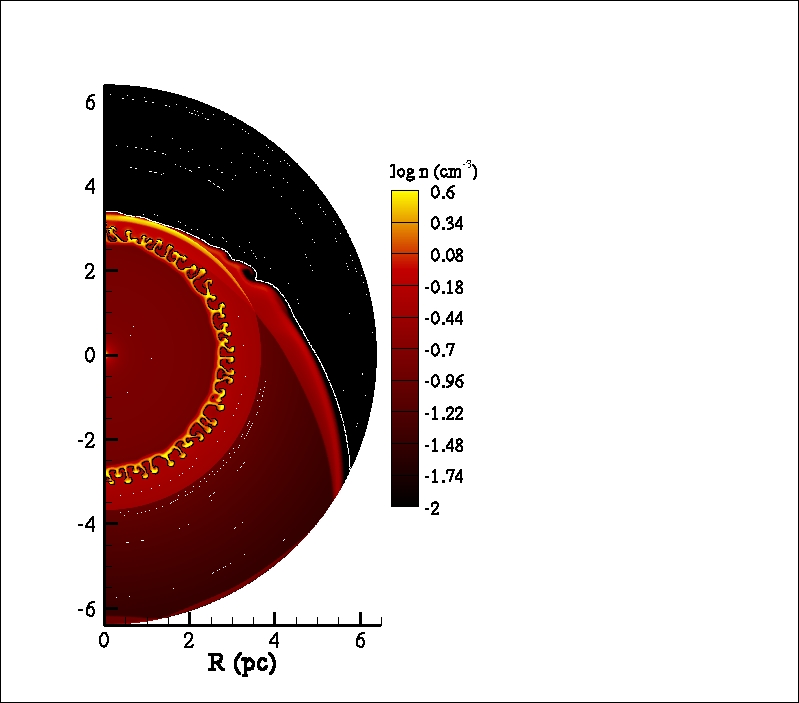} &
\includegraphics[trim=55 25 320 45,clip=true,width=45mm,angle=0]{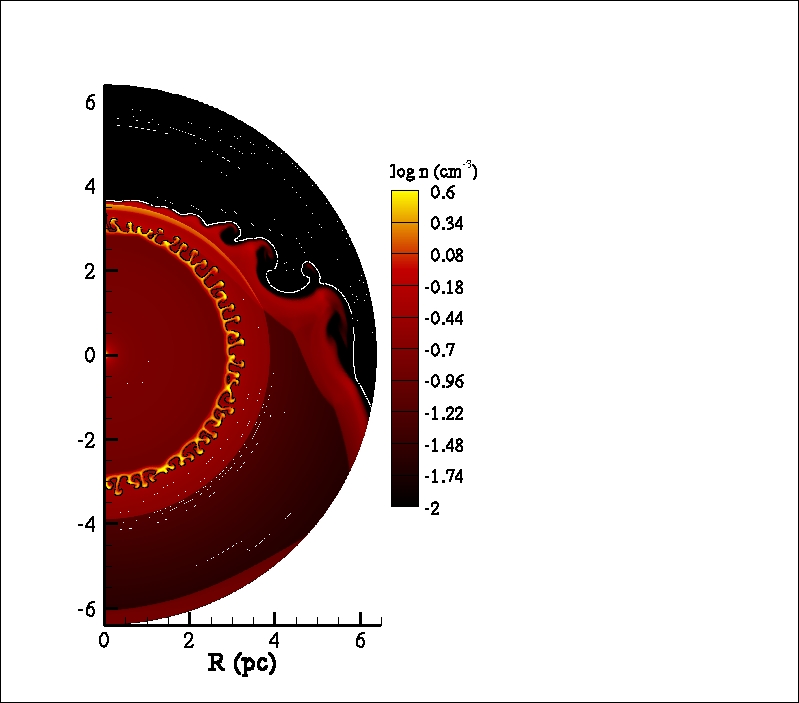} &
\includegraphics[trim=55 25 320 45,clip=true,width=45mm,angle=0]{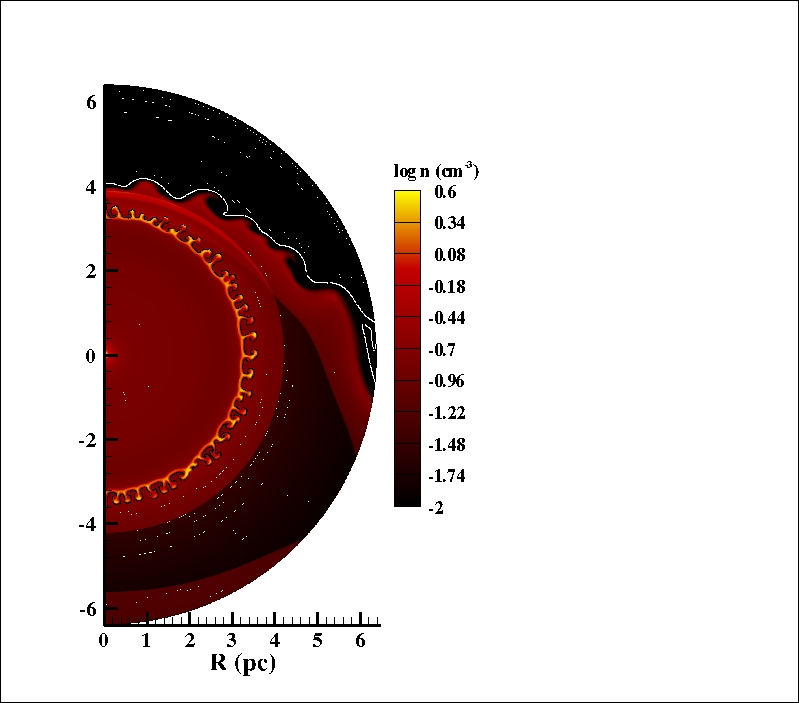}&
\includegraphics[trim=55 25 320 45,clip=true,width=45mm,angle=0]{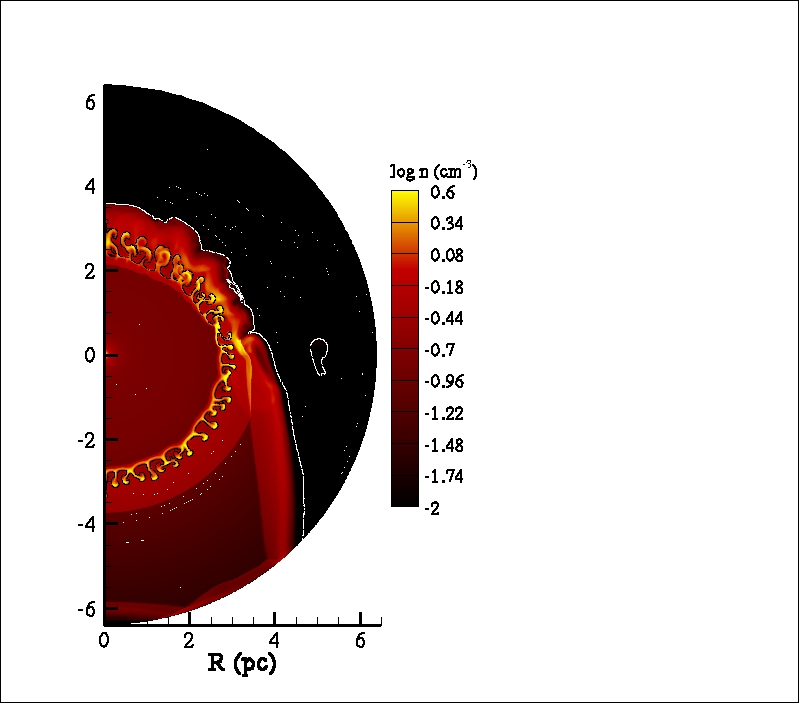} &
\includegraphics[trim=55 25 320 45,clip=true,width=45mm,angle=0]{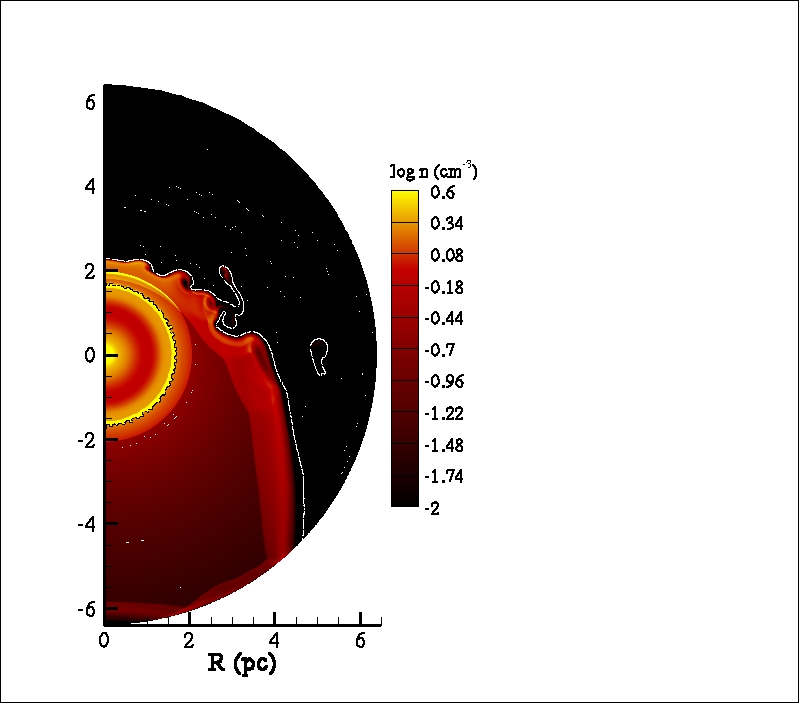} 
\\
\\
\includegraphics[trim=55 25 320 45,clip=true,width=45mm,angle=0]{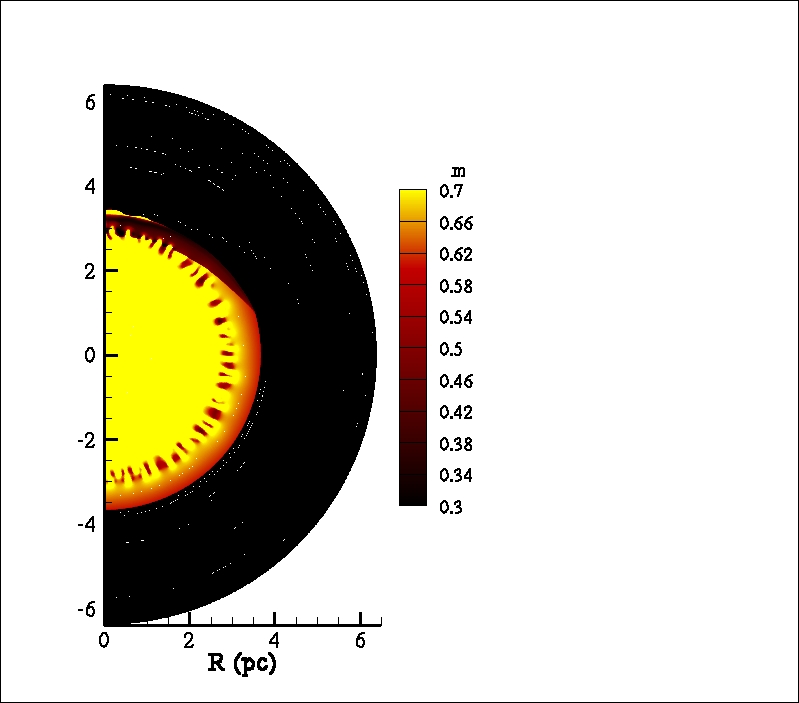} & 
\includegraphics[trim=55 25 320 45,clip=true,width=45mm,angle=0]{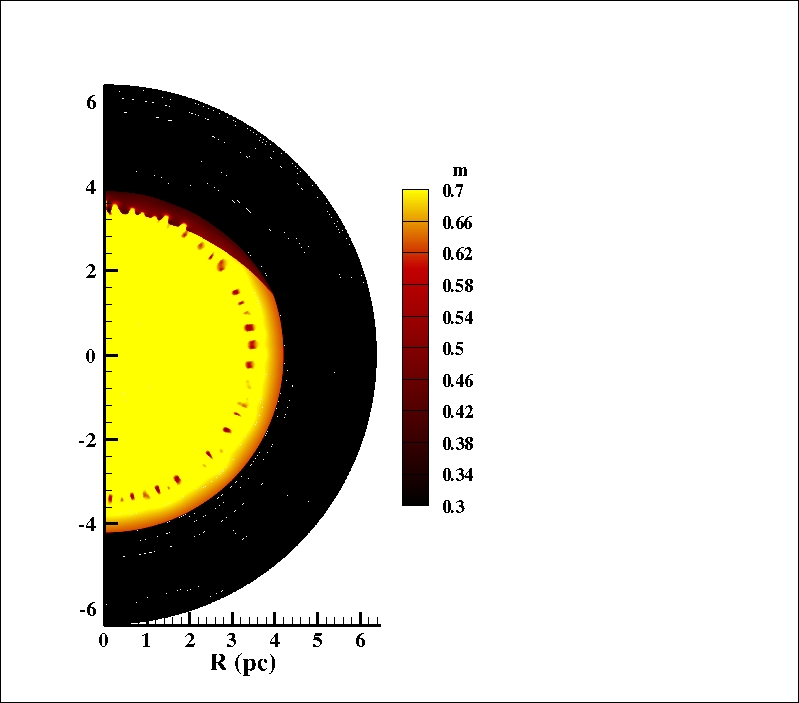} &
\includegraphics[trim=55 25 320 45,clip=true,width=45mm,angle=0]{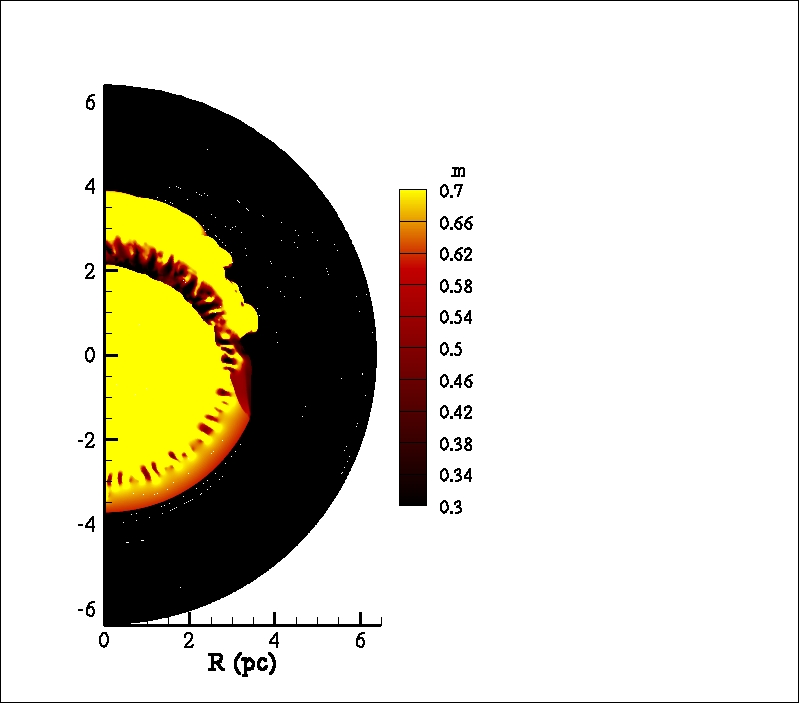} &
\includegraphics[trim=55 25 320 45,clip=true,width=45mm,angle=0]{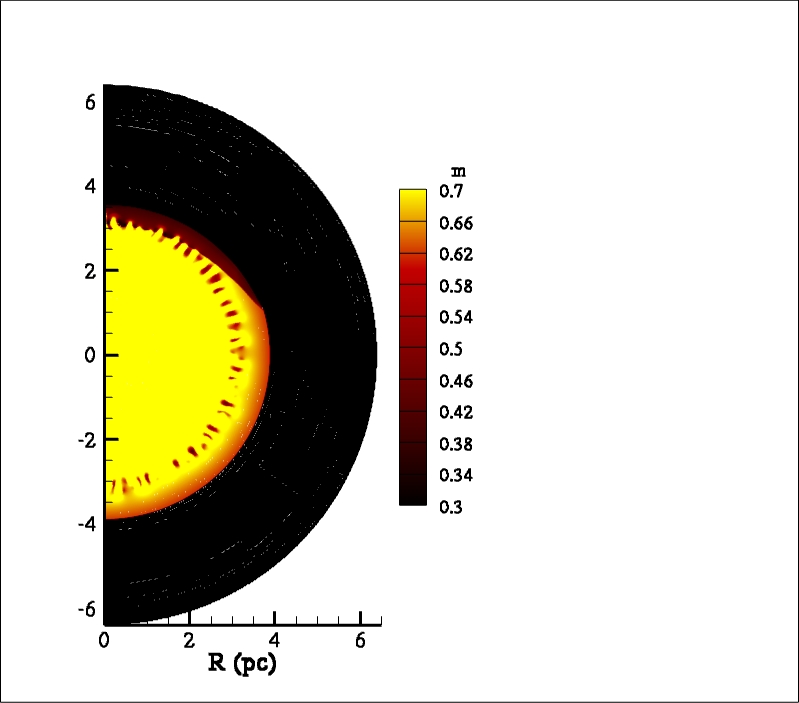}  &
\includegraphics[trim=55 25 320 45,clip=true,width=45mm,angle=0]{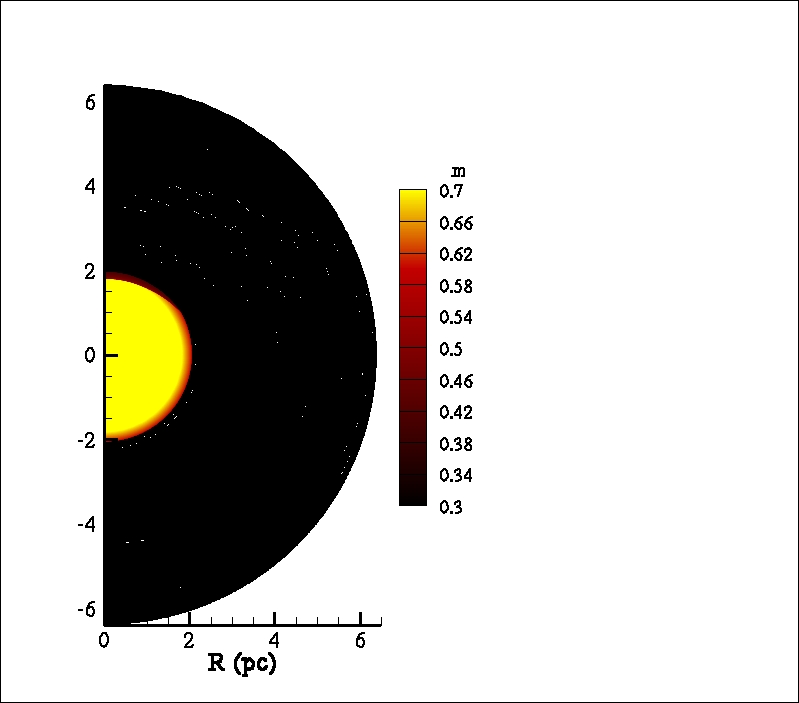}  
\end{array}$
\end{center}
\caption{Comparison of the density and the expansion rate of the five  models, from left to right corresponding to: A, B, C, D, and Dsub. The upper row shows the density of the remnant. The lower panels show the expansion rate. }
\vspace{-17cm}
  \label{SN_comp}
\end{sidewaysfigure*}

\medskip\noindent
{\bf SNR}\\
The location, density, thickness and radius of curvature of the shell all determine the morphology of the SNR upon interaction with the ejecta. 
Figure~\ref{SN_comp} shows the resulting densities and expansion rates of the SNR
at an age of 412~yr, similar to the current age of Kepler.

As we noted in Sect~\ref{sec:snrA}, in Model~A the SNR blast wave reaches slightly beyond the shell in the region around the stagnation point, and accelerates again as it protrudes. The expansion rate of the plasma in the interaction region is consistent with the observed values and it reaches a minimum ($\sim 0.33$) just behind the blast wave. This region is characterized by the highest densities and thus, the emissivity is expected to be also relatively high. 

Apart from the protrusion through the shell, model B yields similar properties.  At the snapshot, which corresponds to the current age of Kepler, almost one third of the remnant interacts with the wind shell. The  radius of the SNR is larger (4~pc compared to 3.8~pc in Model~A), and the  minimum expansion rate is slightly higher due to the lower densities of the wind region and wind shell respectively. 
 
These effects are even stronger for the case of model C. Due to the much lower mass loss rates, the radius of the remnant is 4.1~pc, while the minimum expansion rates tend to the value of 0.4. However, a similar morphology with the other two models has been formed also in this case.

In Model D the asymmetry is  the highest. The interaction is strong because of the proximity and higher mass of the wind bubble, which was designed to accommodate the small distance estimate of 4~kpc to Kepler's SNR. The blast wave reaches well beyond the extent of the shell resulting in a high expansion rate ($\geq 0.7$) in that area. This, together with the high degree of asymmetry, makes this model inconsistent with the observations of Kepler.

In all of these models, the radius of the SNR is fairly consistent with the larger distance estimate ($\sim 6$~kpc) of Kepler's system.
We find that in order to explain the observed morphology in case Kepler is as close by as $4$~kpc, we need a subenergetic explosion. We adopt the parameters of model D for the bubble, so as to have a wind shell relatively close to the star, while decreasing the explosion energy to one fifth of the canonical value in model Dsub.  The radius of the SNR then becomes $\sim 2$~pc, and the asymmetry and expansion rates are consistent with Kepler's SNR. 

From these models, it is clear that the morphology does not depend on fine-tuning of parameters. For a reasonable range of values for all of the parameters, the expansion rate, size, density, and morphology of Kepler's SNR are well reproduced.  The different models result in some variation in the blast wave position and locations of maximum density and minimum expansion rates.

\section{Summary and discussion}\label{Sect:Discuss.}

\subsection{Comparison of Kepler's SNR with the models}
We have argued above 
that the kinematics, morphology and abundances in the northern region of
the Type Ia SNR Kepler are best explained by a model, in which the 
progenitor system was a symbiotic binary with a
WD accreting mass from the slow (10--20~\kms) wind from an AGB donor star.
The accretion efficiency is moderate and most likely of the order of 10-50\%, which characterizes the wind, or the wind-RLOF mass transfer.  This allows for
sufficient wind luminosity, needed for creating a shell, 
while, at the same time, it allows the WD to accrete at a sufficient rate
to reach the 1.38~\msun\ within the lifetime of the AGB phase of the donor star. 
The initial mass of the AGB star must have been 4--5~\msun,
in order to explain the enhanced nitrogen abundances in the northern part of the
SNR. 

The wind emanating from the system creates a bubble, 
which, due to  supersonic systemic motion of the system (250 \kms), is surrounded
by a one-sided shell, created by the bow shock. This explains both the morphology of Kepler's SNR, and
its height above the Galactic plane of 590$d_5$~pc.
In this sense our model is reminiscent of the stellar runaway model
for Kepler proposed by \citet{1988LNP...316...81B} and \cite{1992ApJ...400..222B}, but applied to a Type Ia progenitor system.

The parameters of the model are well constrained,
as the distance between the progenitor system and the wind shell
depends on a combination of the systemic velocity, wind velocity, mass-loss
rate and ISM density (Eq.~\ref{eq:stag}, Fig.~\ref{fig:stagnation}). 
At the same time, the mass-loss rate
should be high enough to allow for enough accretion onto the WD within the
AGB phase, while the total mass that is lost from the system is limited by
the envelope mass of the donor star, $\sim 3-4$~\msun. A final constraint is
that the wind velocity should be high enough to allow for the creation of
the shell within the lifetime of the AGB phase (Eq.~\ref{eq:tflow}).

We tested our model with several hydrodynamic simulations of
both the creation of the stellar wind bubble and the subsequent
SNR evolution, varying the systemic velocity, the
wind velocity, and mass loss rate.
Since there is some uncertainty about the distance to Kepler's SNR,
and hence about the size of the wind shell, we included also two models (model D and Dsub )
resulting in a stagnation point at 2~pc instead of 3~pc.

All models reproduce the overall characteristics of
Kepler's SNR; they have one-sided shells, and the expansion parameter
of the SNR inside this dense shell drops to $m=0.3-0.4$, in agreement
with observations \citep{1988ApJ...330..254D,2008ApJ...689..231V,katsuda08}. An exception is 
model D. This model has the stagnation point of the wind at 2~pc,
relevant for a distance of Kepler's SNR of $\sim 4$~kpc.
The shock of the SNR has moved completely through the dense shell after
400~yr, and has an expansion parameter of $m>0.7$, inconsistent
with the observations. In this case, in order to match the model with the 
observations, the SN explosion energy has to be $E=0.2\times 10^{51}$~erg 
(model Dsub),
which was also suggested by \citet{2008ApJ...689..231V}.  Since normal SNe Ia
have explosion energies in excess of $10^{51}$~erg \citep{2007ApJ...662..487W}
and given the high iron content and historical light curve \citep{baade43},
SN 1604 must 
have been a fairly normal Type Ia.
Our models therefore favor the larger distance estimate of Kepler's SNR,
$d \approx 6$~kpc.

Although the overall morphology and kinematics of Kepler's SNR can 
be reproduced, there are some interesting differences between the models.
Most notable is the fact that in model A the blast wave has just
emerged out of the shell near the stagnation point.
The question is now
whether this is true for Kepler's SNR as well. This question is related to
the orientation axis of Kepler's SNR with respect to that of our model.
Examining the available literature and X-ray data we think that a good
case can be made that the north is the projected direction of
systemic motion of the progenitor system, and that the
shock wave in Kepler has indeed just emerged out of the wind-created shell.

The first reason is that the nitrogen rich knots are found mostly
in the northwest, and also in the northeast, but in the north
they are almost absent \citep[Fig.~1 and 2 in][]{1991ApJ...366..484B}. Non-radiative
shocks associated with fast shocks in lower density gas
are more prevalent due north \citep{2005AdSpR..35.1027S}. 
Similarly, more  4-6 keV X-ray continuum emission
associated with the
shock front (blue in Fig.~\ref{fig:chandra})
appears to be present in the north than in the northeast and 
northwest. Most of the X-ray continuum emission in Kepler's SNR
seems to be synchrotron radiation \citep{2007ApJ...668L.135R,2008ApJ...689..231V}, 
which requires fast shock velocities $V_s \gtrsim 2500$~\kms 
\citep{aharonian99,2007A&A...465..695Z}. This suggests that
the shock has picked
up speed in the northern region. 

 Additionally, the $\rm H_\alpha$ images show a much thinner region of the emission due north compared both to the rest of the `shell' and with the X-ray images. This could well be due to the fact that in the ISM hydrogen is completely ionized and no Balmer shocks can be expected beyond the shell. This would be an additional argument in favor of the shock having penetrated the shell. However,in order  to make a firm conclusion based on this feature a more detailed comparison between the X-ray and $\rm H_\alpha$ images is required.
Although, in our view, model A  seems to reproduce Kepler's SNR best,
slight adjustments of the parameters of model B and C may
also result in a blast wave extending just beyond the shell.

Finally, in our comparison of the models with Kepler's SNR
we would like to point out that the SNR shell in the models in the
north is thinner than in the south. This brings also the
contact discontinuity of ejecta and swept up matter closer to the shock
front. This, indeed, seems to be the case in Kepler's SNR because
the emissivity profiles corresponding to different ejecta elements
show that in the northern region
there is less stratification of the ejecta and the ejecta are closer
to the forward shock than in the southern region \citep{cassam04}.
A similar situation seems to arise in Tycho's SNR, where the proximity
of the ejecta to the shock front has been attributed to a change
in the effective equation of state as a result of efficient cosmic
ray acceleration \citep{warren05}. Our simulations, however, can explain this, at least partially, based on pure hydrodynamics alone \citep[c.f][]{Kosenko2010}.

\subsection{Implication of the model for the progenitor of SN1604}
The symbiotic binary model that we advocate here for Kepler has some
interesting consequences for the current state of the donor star.
So far, the identification of a donor star of a historical supernova has only been
claimed for Tycho/SN 1572 \citep{ruiz04}, which was mainly based on its large proper motion. However, this result is
still controversial,  since the large proper motion of the candidate star is not  accompanied by a high spin velocity as would be expected for close Roche-lobe overflowing binary progenitors \citep{kerzendorf09}. In our model no such high orbital and spin velocities are expected for the donor star because we require a wide symbiotic binary. It should nonetheless maintain the high systemic velocity of the progenitor system of $\sim 250$ ~\kms, something that constitutes a clear signature for its identification.

The donor star of Kepler is expected to be, according to our model,
an  AGB star with initial mass of 4-5~\msun, which has lost almost its entire envelope due to the mass transfer  and the subsequent collision with the blast wave of the supernova \citep{marietta00}. After the collision and the mass strip effect, the donor star will expand to re-establish its thermal equilibrium. Stars with convective envelopes are characterized by short thermal timescales, which means that  the donor star has most likely already re-expanded to its original size. However, further investigation is needed before a firm conclusion can be be drawn
about the present-day characteristics of the donor star, as the expansion timescale is determined by the mass of the stripped layers and the energy deposited at the remaining layers of the donor star due to the collision with the SNR  \citep{2003astro.ph..3660P}. 

Three possible cases can be distinguised: {\em i)} The donor star has re-established its thermal equlibrium: In this case the donor star should be an evolved star with a bolometric
luminosity of an AGB star with initial mass of 4-5~\msun. The reason for this is that giants'
luminosity is mainly determined by the  mass of the core, which remained unaffected by the collision. We, therefore, expect an absolute magnitude
of $M_V=-4.5\pm 0.5$, which at a distance of 6~kpc and $A_V=2.8$ 
\citep{2007ApJ...668L.135R} implies an apparent magnitude of $m_V=12.0\pm 0.5$. Due to the mass loss, the surface gravity of the donor should be much lower that this of a 4-5~\msun\ AGB star and perhaps the remaining envelope
has picked up ejecta material, and has elevated abundances of iron
and intermediate mass elements (Si, S, Ar, Ca).  {\em ii)} The donor star is still in the expanding phase: In this case the star should be underluminous and cooler, and may have the appearence of a later type star. {\em iii)} The donor star has lost its entire envelope: In this case the donor's remnant should be a massive ($\sim 1$ \msun) , young CO white dwarf.

\subsection{Is the progenitor system of SN1604 typical for Type Ia supernovae?}

The model outlined here is specifically intended to explain many characteristics
of Kepler's SNR. Nevertheless, given the many problems surrounding conventional
models for Type Ia progenitors (see the Introduction), it is worthwhile
to discuss whether our model could be more widely applicable  than just to this historical SN.

 The progenitor system in our model is a symbiotic binary, where the WD accretes mass from its AGB companion through non-conservative mass transfer.
This progenitor system is not a conventional progenitor in the rich literature of Type Ia SNe. The reason is that RLOF of these binaries leads to unstable mass transfer, while wind accretion is characterized by low accretion efficiencies ($\leq 10 \%$), making it difficult to result in Type Ia events. However, these progenitor systems are currently still far from being understood.  The Bondi-Hoyle model that is used to describe the wind accretion is rather simplified and not applicable to slow winds \citep{2004NewAR..48..843E}, while the much more efficient wind-RLOF mass transfer  has not yet been studied in terms of population synthesis. Observations already  seem to reveal  
symbiotic systems with AGB donor stars,  which, despite their large separation, show evidence of direct mass transfer  and substantial hydrogen burning at the WD's surface (Mira AB: \citet{2004RMxAC..20...92K}; V 407 Cyg: Mikolajewska 2010, private communication). Furthermore, an non-negligible number of SNRs, identified as Type Ia,  display spectroscopical evidence of strong interaction between the ejecta and dense CSM such as SN 2002ic \citep{Hamuy2003} and SN 2005gj \citep{Aldering2006}  and SN 2006X \citep{Patat2007}. In addition, \citet{Borkowski2006} showed that the highly centered concentrated Fe emission of the two  Type Ia SNe in LMC DEM L238 and DEM L249 can best be explained assuming that the ejecta interacts with a non-uniform CSM.  All these observations point towards a SD scenario, where the circumstellar matter was formed by the slow wind of a giant, or an AGB donor star. Thus, non-conservative mass transfer of 
 symbiotic binaries  should be considered as a viable path towards Type Ia SNe.

The  generally favored SD model of a Roche-lobe overflowing donor star and an accreting, nuclear burning WD has two major problems with observations of Type Ia SNRs. First, the presence of neutral hydrogen around young Type Ia SNRs
is at odds with the presence of large ($\sim 30$~pc) HII regions
around supersoft sources \citep{ghavamian03}. Second, there seems to be a lack of the observed X-ray emission in the elliptical and spiral galaxies (\citet{Gilfanov2010,DiStefano2010}, respectively),  which is expected from the steady hydrogen fusion on the WD surfaces \citep[however see][]{Hachisu2010}.   The symbiotic binary model for Kepler does not present either of these problems.

 The accreting WD in our model is enshrouded by the dense
wind from the donor. The column density towards the WD in such a case
is given by:
\begin{eqnarray}
N_{\rm H} &\approx &\int_{R_{in}}^\infty \frac{\dot{M}_w}{4\pi r^2 u_w}
\frac{1}{1.4m_{\rm p}} dr = \\ \nonumber
\\\nonumber
 &&3.6\times 10^{22}
\Bigl(\frac{\dot{M}_w}{10^{-5}{\rm M}_\odot\,{\rm yr}^{-1}}\Bigr)
\Bigl(\frac{u_w}{20\,{\rm km\,s}^{-1}}
\frac{R_{in}}{3\cdot 10^{14}\, {\rm cm}}\Bigr)^{-1}
\ {\rm cm}^{-2},\nonumber
\end{eqnarray}
with $R_{in}$ corresponding to the separation of donor star and WD.
Even for a mass-loss rate of 
$\dot{M}_w =3\times 10^{-6}$~M$_\odot$\,yr$^{-1}$, 
we still have $N_{\rm H}= 10^{22}$~cm$^{-2}$, sufficient to attenuate
the UV/X-ray flux by a factor $10^{4}$
\citep[based on the absorption model of][]{wilms00}.
This does not only explain
the lack of extended HII regions around young, nearby Type Ia SNRs,
but it also suggests that the progenitors of SNe Ia may not
necessarily be detectable in X-rays.

Note that the mass-loss rates of the models in the range of 
$10^{-6}-10^{-5}$~\msun\,yr$^{-1}$ are lower than or near the upper limits
inferred by the lack of narrow H and He emission lines and X-ray emission from
SNe Ia \citep[][respectively]{mattila05,hughes07}, but much higher than the strictest upper limit to mass loss rates set by radio observations, which gives $\dot{M}  \sim 3 \times 10^{-8}~(u_w/10~ \rm{km~s^{-1}})$~M$_{\odot}$yr$^{-1}$ \citep{Panagia2006}.

The last limit poses a challenge  to our model. However, even if the lack of radio flux in Type Ia SNRs puts clear constraints to the density of the ambient medium, the numerical lower limit estimated by \citet{Panagia2006} is debatable as it relies on semi-empirical models based on the radio emissivity of SNe Ib/c. In addition, the solution suggested by \citet{WoodVasey2006}, where successive nova explosions, just before the final Type Ia event, form an evacuated region around the explosion center, is also possibly applicable for the case of Kepler's SNR.

 An extra challenge to our model is related to the size of the donor star. \citet{Hayden2010} and \citet{Tucker2011}  analyzed the early-time lightcurves of several hundreds SNIa in the optical   and UV  band and they found no signatures from the blast wave interaction with the donor star, predicted by \citet{kasen10}. This result constrains evolved stars with large radii as possible candidates of WD companions in the SD Type Ia regime.  Nevertheless, two reasons can possibly exclude our suggested model from this category. Firstly, even if the radius of a  $4-5$ \msun~AGB star is large, a non-conservative mass transfer demands a  wide binary system. Thus, the expected fraction of SNIa, which shows the signature of the collision at their lightcurve are much smaller (by a factor of ten) than the estimated one for the  Roche-lobe overflowing  systems \citep[see eq. 9][]{kasen10}. Secondly, most of the enhanced optical/UV 
emission caused by the interaction
of the ejecta with the donor star will in our model be absorbed by the dense AGB wind.
However, more quantitative conclusions regarding the light-curve for wind accreting Type Ia progenitors should be based on detailed hydrodynamical and radiative
modeling of the immediate aftermath of the Type Ia explosion, similarly to what 
\citet{kasen10} has done for Roche-lobe overflowing binary systems.

{A second implication of our proposed progenitor model is that the high column densities suggest significant
absorption during the first few days following the explosion, when
the supernova blast is carving itself out of the denser parts of the 
wind. Assuming again  $\dot{M}_w =3\times 10^{-6}$~M$_\odot$\,yr$^{-1}$,
$u_w =20$~\kms, and a supernova shock velocity of 25,000~\kms,
then the optical light passes through a column of at least $N_{\rm H}\approx
10^{22}$~cm$^{-2}$ one day after the explosion, $4\times 10^{21}$~cm$^{-2}$
after four days, and $2\times 10^{21}$~cm$^{-2}$ around 8 days after
explosion, corresponding roughly with the time around maximum luminosity. 
Using the Galactic value for converting $N_{\rm H}$ into $A_V$ 
\citep{predehl95x}, around peak luminosity the absorption is expected to be
 roughly $A_V\approx 1$. Therefore, the supernova should have had an appreciable
intrinsic reddening the first few days, 
if it originates from a symbiotic system. Interestingly,
early light curve data of SN1994D \citep{patat96} and SN1996X \citep{salvo01},
indeed, show that before maximum light the supernova 
initially becomes bluer and after maximum light becomes
redder again. This might be expected if the blue atmosphere is initially
reddened by the wind bubble, and, once it has emerged out of the wind bubble,
it becomes redder again, but then caused by 
the adiabatic cooling of the ejecta.
In this light it is also interesting that, when SN1604 was first observed,
its color was described to be ``like Mars'', and only on day eight
it was described as ``like Jupiter'' \citep{baade43}. Note that
both Jupiter and Mars were in conjunction at the time, and SN 1604 appeared
very nearby these planets, so atmospheric extinction was very similar for
all three objects. This conjunction was in fact the reason that SN 1604
was detected so early on \citep{2003LNP...598....7G}.

We like to end this section by pointing out that the one-sided shell, formed close to the explosion center of Type Ia SNe, could be rather common.  The properties of the shell depend on the details of the
wind parameters, the systemic velocity of the progenitor system and the local density of the ISM. Thus, even if the systemic velocity adopted here for the case of Kepler's SNR is quite extreme for low mass binary systems, similarly sized shells can be formed considering a progenitor system that has a much lower velocity but moves through a denser ISM.  Interestingly, several
Type Ia SNRs  both show  evidence for an axis of symmetry,
and seem to encounter a density enhancement in one
direction. Two examples are Tycho's SNR (SN1572), which has 
a lower expansion velocity in the east \citep{katsuda10}, and SN\,1006,
which is very symmetric in the northeast/southwest direction, but has
a higher density in the northwest.

We offer these examples as tentative evidence that Kepler may not be as unique
as we made it out to be in the Introduction. Rather, the one-sided shell
in Kepler may just be more apparent than in other SNRs.

\section{Conclusions}\label{sec:conclusions}
In this paper we have presented evidence that the observational 
characteristics
of the Type Ia SNR  Kepler/SN1604 are best explained by a moving
progenitor
system, in which the donor star forms a wind blown bubble, while at the
same time the WD accretes part of the wind, thereby reaching 1.38~\msun\
within $\sim 0.4-0.7$~Myr. Due to
the systemic velocity the wind bubble is surrounded by an asymmetric 
shell,
reaching its highest densities in the direction of the motion of the
progenitor.
In order to explain the nitrogen abundance
in the north of Kepler, the donor star probably had a main sequence 
mass of
4-5\msun, which are known to have nitrogen rich envelopes in the AGB 
phase.

We used hydrodynamic simulations to show that the wind properties
of the system can explain the observational
characteristics of Kepler's SNR,
namely a one-sided shell with which the SNR is interacting,
and a slower expansion velocity in that region.

Our simulations show that in the direction of the progenitor's velocity
the SNR blast wave may just have completely penetrated the wind blown
shell, or still be inside it. Based on the possible presence of
X-ray synchrotron radiation in the north of Kepler, and the lack of it
in adjacent regions, and the absence of nitrogen rich knots in that 
direction,
we argue that in Kepler part of the blast wave may indeed have 
penetrated
all the way through the shell.  On the other hand, the presence of Balmer dominated shocks indicate that at least part of the blast wave must still be within the cool, neutral, wind shell.

Our results also show that the distance to Kepler's SNR is likely
to be around 6~kpc, and not 4~kpc \citep{2005AdSpR..35.1027S}. This 
larger
distance estimate is in agreement with recent values based on the absence
of gamma-ray emission \citep{2006A&A...452..217B,2008A&A...488..219A}, and the kinematics
of Kepler's SNR \citep{2008ApJ...689..231V}. However, a distance of 
4~kpc can be
reconciled with the simulations if the explosion energy was
$2\times 10^{50}$~erg. This would make SN 1604 a subenergetic explosion,
which seems unlikely given its historical light curve and the
copious amount of Fe present in the SNR.
According to our scenario, the donor star should still be present within the SNR. It is likely
an evolved star with most of its envelope mass stripped.

We have argued that our model may be more broadly applicable to
Type Ia supernovae and their progenitors. The presence of a dense
wind close to the supernova should result in considerable intrinsic
reddening of
the supernova during the first few days after the explosion.
For the pre-supernova appearance of the system, the large
absorption column blocks out most of the X-rays from the
accreting WD. These systems would, therefore, be classified as 
symbiotic binaries,
and not as supersoft sources. This also explains the presence of
neutral hydrogen near young Type Ia SNRs, which cannot
be present around supersoft sources due to their high X-ray/UV flux
\citep{ghavamian03}.

Another consequence of the model, if it is more widely applicable to
Type Ia progenitors, is that many SNRs should be interacting with one
sided shells. For Tycho's SNR this could  explain the low expansion rate
in the east, and even, at least partially,
the proximity of the ejecta close to the forward
shock, an observational characteristic usually attributed to the 
presence
of cosmic rays \citep{warren05}. 

Kepler's SNR requires a rather narrowly defined model. So major question remains whether
SN 1604 was a very special event with an unusuale progenitor history, or on the other should be considered to be a key object 
for a better understanding of Type Ia supernovae in general.
\\
\\
\textit{Acknowledgments.}  We are  grateful to  Arend-Jan Poelarend, Onno Pols and Selma de Mink for many helpful  discussions on the topics of stellar and binary stars evolution. Also, we thank Rony Keppens for providing us with the AMRVAC code. 
We thank Frank Verbunt and Rony Keppens for their
helpful suggestions that helped us  improve the manuscript.
This work is supported by  Vidi grant from the Netherlands Organisation for Scientific Research (NWO).

\bibliography{article1} 

\end{document}